\documentclass[final,5p,times,twocolumn,letterpaper]{elsarticle}

\usepackage{amssymb}

  \usepackage{amssymb}
  \usepackage{latexsym}
  \usepackage{amsfonts}
  \usepackage{amsthm}
  \usepackage{amsmath}
  \usepackage{graphics}
  \usepackage{setspace}
  \usepackage{graphicx}
  \usepackage{color}
  \usepackage{float}
  \usepackage{wrapfig}
  \usepackage{color}
  \usepackage{rotating}
  \usepackage{dblfloatfix}
  \usepackage[hyphens]{url}
  \usepackage[table]{xcolor}
  \usepackage{multirow}

  \usepackage{graphics}
  \usepackage{setspace}
  
  \usepackage{graphicx}
  \usepackage{epstopdf}  
  \usepackage{color}
  \usepackage{float}
  \usepackage{wrapfig}
  \usepackage{color}
  
  \usepackage{graphicx}
  \usepackage{caption}
  \usepackage{subcaption}
  \usepackage{setspace}
  
  \definecolor{blue}{rgb}{0,0,0}

\usepackage{lipsum}
 \usepackage{hyperref}

 \hypersetup{colorlinks=false}

  \usepackage{algorithm}
  \usepackage{algorithmic}
  \usepackage{morefloats}
   \usepackage{textcomp}

   \usepackage{picins}

\newenvironment{noindlist}
 {\begin{list}{\labelitemi}{\leftmargin=1.2em \itemindent=-.5em}}
 {\end{list}}

\journal{Knowledge-Based Systems}

\begin{document}

\begin{frontmatter}

\author[rvt]{Charith Perera\corref{cor1}}
\ead{charith.perera@ieee.org}
\author[rvt3]{Athanasios V. Vasilakos}
\ead{vasilako@ath.forthnet.gr}

\fntext[fn1]{An earlier version of this work has been published in the Proceedings of the 9th International Conference on Semantics, Knowledge \& Grids (SKG).}

\address[rvt]{Centre for Research in Computing, The Open University, Milton Keynes, UK}
\address[rvt3]{Dept of Computer Science, Electrical and Space Engineering, Lulea University of Technology, Lulea, Sweden}

\title{A Knowledge-Based Resource Discovery for Internet of Things \vspace{-5pt}}


\author{}

\address{}

\begin{abstract}
\textcolor{black}{In the sensing as a service paradigm, Internet of Things (IoT) Middleware platforms allow data consumers to retrieve the data they want without knowing the underlying technical details of IoT resources (i.e. sensors and  data processing components). However, configuring  an IoT middleware platform and retrieving data is a significant challenge for data consumers as it requires both technical knowledge and domain expertise. In this paper, we propose a knowledge driven approach called Context Aware Sensor Configuration Model (CASCOM) to simplify the process of configuring IoT middleware platforms, so the data consumers, specifically non-technical personnel, can easily retrieve the data they required. In this paper, we demonstrate how IoT resources can be described using semantics in such away that they can  later be used to compose service work-flows. Such automated semantic-knowledge-based IoT resource composition approach advances the current research. We demonstrate the feasibility and the usability of our approach through a prototype implementation based on an IoT middleware called Global Sensor Networks (GSN), though our model can be generalized to any other middleware platform.}

\end{abstract}


\end{frontmatter}

\section{Introduction}
\label{sec:Introduction}

The Internet of Things (IoT) \cite{P003} envisions connecting billions of smart devices to the Internet. It provides a networked infrastructure that enables things to be connected anytime, anyplace, with anything and anyone, ideally using any path, any network and any service \cite{P029}. These smart devices should be smoothly integrated within Future Internet (FI) service delivery models such as sensing as a service.  The \textit{things}\footnote{We use terms \textit{objects}, \textit{things}, \textit{smart objects}, \textit{devices}, \textit{nodes} to give the same meaning as they are frequently used in IoT literature interchangeably.} in IoT are accompanied with sensors and actuators. It is estimated that there are about 1.5 billion Internet-enabled PCs and over 1 billion Internet-enabled mobile phones today. By 2020, there will be 50 to 100 billion devices connected to the Internet \cite{P029}. Since these smart devices comprise sensors, it is evident that there would be many sensors deployed around us in the future. Even today, sensors are used in many domains such as agriculture, environmental monitoring, and manufacturing \cite{ZMP007}. 

In order to analyse and understand a given phenomenon extensively, data generated from appropriate sensors needs to be fed into more sophisticated data analysis applications. These applications are designed to produce certain results once they are given required sensor data as inputs. IoT middleware solutions simplify the retrieval of data from sensors for these applications  by acting as a mediator between the hardware layer and the application layer. In order to perform these bindings,  middleware solutions need to be configured  depending on the context information and user requirements. Our objective is to automate and simplify the configuration of  IoT middleware platforms and improve their usability so both IT experts and non-IT experts can use them efficiently and effectively.

There are several characteristics we have identified as important for  developing a model for IoT that provisions sensing as a service by formulating and composing multiple  types of sensor as well as different filtering, fusing, and reasoning mechanisms together on-demand. The core features of the proposed model  are as follows:

\begin{itemize}
\item \textit{Autonomic:} The model should support the dynamic composition
 of  internet-connected objects, in response to dynamically defined end-users' requests. To this end, we have incorporated semantic  knowledge \cite{P071}, along with automated reasoning algorithms for orchestrating sensors, and data processing mechanisms \cite{ZMP007}, according to the  data consumer requests. 

\item \textit{Utility Based:} The proposed model should deliver services according to a utility computing model \cite{Buyya2, P498}. It should offer sensing capability as a service\cite{ZMJ002} over dynamically created and configured solutions\footnote{Solution is a combination of sensors and data processing components that can be composed together in order to satisfy a user requirement.} that are custom generated for each consumer request. Sensor data consumers (users) should be allowed to make the decisions on the characteristic of the solution (e.g. accuracy, reliability, latency and so on). Orchestration of IoT resources (i.e sensors and data processing components) in the cloud environment at runtime is an important functionality \cite{IOTABOOK}.  The dynamism implies the capability of adapting to resources changes in volatile environments where sensors and data processing components may be added or removed from the system over time. This means that new solutions will be able to compose together over time due availability of new resources.

\item \textit{Scalability and Flexibility:} The proposed solution should be flexible so data processing components and sensors can be added over time \cite{Teixeira}. Further, the proposed solution should be scalable so any number of IoT resources can be supported. Such ability increases the types of consumer requests that can be fulfilled. Further, it increases the number of different solutions that can be formulated to accommodate a single  request. Possibility of creating multiple different solutions will increase the choice and control consumer have. Finally, the proposed model and its algorithms should be independent from the data  (i.e. descriptions of IoT resources) so adding a new resource does not require changes to be made into the system.

\item \textit{Ease of use / reduced learning curve:} One of the primary goals of an IoT middleware is to enable users to retrieve data quickly without dealing with complex hardware or software level configurations. It is important to make all the process simplified so a non-technical personal (e.g, biologist) can use these IoT middleware platforms to collect the data they need with minimum effort.
\end{itemize}

\subsection{Motivation}
\label{sec:Motivation}

Over the last few years, we have seen more and more IoT middleware platforms making their way in the marketplace. Large number of sensors are expected to connect to these middleware platforms. Further,  variety of different IoT applications are expected to be built on top of these middleware platforms. These IoT applications have different types of algorithms that analyse data built into them. These algorithms are nothing but some type black boxes that take specific type of inputs and generate specific type of outputs. One of the main responsibilities of an IoT middleware is to hide and abstract the connectivity and communication details of sensors and support the users to retrieve the data streams they required to be fed into their application easily and quickly. 

Data processing components can also be used to build these required data streams as we  later discuss in this paper. Currently, it is difficult to configure IoT middleware platforms in a way that they produces a certain data stream that is required by an IoT application. The challenges  are discussed in Section \ref{sec:Background}. To make IoT middleware configuration easier, we propose a knowledge driven approach called Context Aware Sensor Configuration Model (CASCOM) to simplify the process of configuring IoT middleware platforms, so the data consumers, specially non-technical personnel, can easily retrieve the data they required.

\subsection{Main Contributions}
\label{sec:Main Contributions}

The contributions of our paper are as follows:

\begin{itemize}
\item We propose a IoT configuration model called CASCOM to enrich  existing IoT middleware platforms. This model helps non-IT experts to configure sensors and data processing components with less effort.

\item CASCOM is completely driven by semantically enriched IoT resource descriptions at the back end. Therefore, new sensors and data processing components can be added at any time. No changes are required in the application from an algorithmic perspective.

\item CASCOM provides an easy way to construct the data streams required by the data consumers by selecting questions and answering them.

\item CASCOM automatically highlights to users on potential secondary context information that can be derived from existing primary context.

\item Finally, CASCOM informs the users regarding potential sensor and data processing components that can be added to the system in order to enhance the ability to serve the user requests. 

\item \textcolor{black}{CASCOM uses ontologies to model  semantics where three ontologies capture the relevant knowledge collectively. The usage of ontologies help to automated composition and reasoning process. More importantly, modelling new knowledge is very easy due to the adoption of ontology based knowledge modelling technique. Specifically, we employ two existing ontologies, namely SCO \cite{P612}, and SSN ontology \cite{P626}} and developed our own ontology called QA+TDO. New knowledge can be added to these existing models easily using the proposed tool.

\end{itemize}

\begin{figure*}[b!]
 \centering
 \includegraphics[scale=.65]{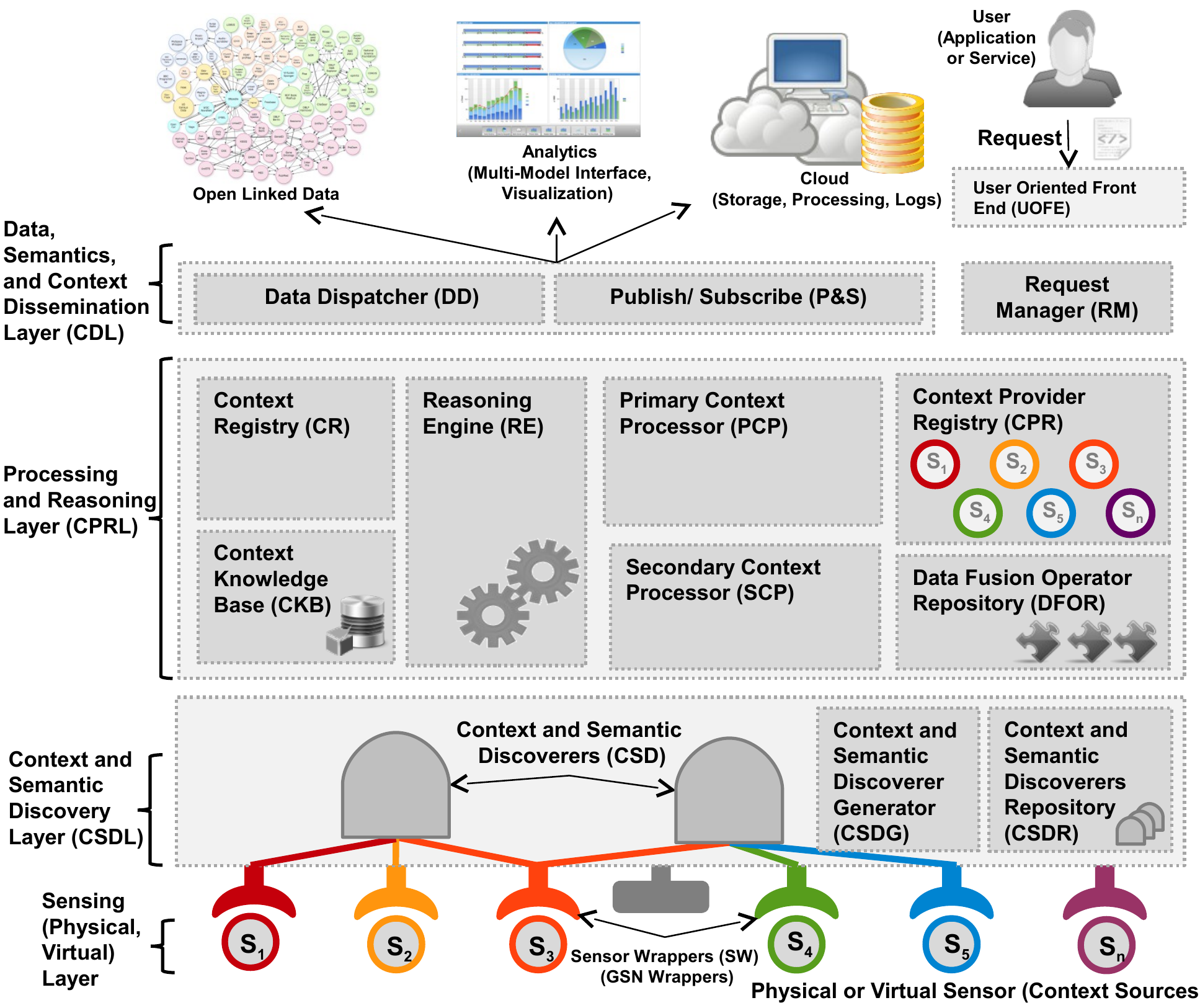}
 \caption{\textcolor{black}{Internet of Things Reference Architecture. Our proposed model, CASCOM, fits within the Reasoning Engine (RE) block of the architecture. The details of this architecture is discussed in detail in \cite{ZMP004}}}
 \label{Figure:Reference_Architecture}	
\end{figure*}

We explain all the above mentioned contributions in detail throughout the paper. The rest of this paper is structured as follows. Section \ref{sec:Background} presents the background and related work. First, we  briefly introduce an IoT reference architecture and its characteristics. Then, we explain where our proposed model fits in such an architecture. Later, we review some related work and compare them with our own to highlight the similarities and differences. The research challenges are discussed in Section \ref{sec:Problem_Analysis}.  We have used a real world use-case scenario from the agriculture domain to explain the research problem in detail. Our research question is \textit{`How to develop a model that allows data consumers (i.e, non-IT and IT experts) to configure IoT middleware platforms by discovering and composing IoT resources (i.e, sensors and data processing components) effortlessly?'}. Once the configuration is completed, the IoT middleware platform should produce the data streams that the data consumers have requested. Data consumers can use these data streams to achieve their own objectives.

Subsequently, we explain the importance of resource discovery and composition in IoT domain and why it need to be knowledge-driven. Architectural designs are presented in Section \ref{sec:Architectural_Design}. We propose our solution, CASCOM, which consists of six phases where each phase is explained in detail with relevant algorithms and examples. Section \ref{sec:Implementation} presents the implementation and experimentation details. We evaluate the proposed model from both computational (i.e. storage requirements, data model loading time, query time) and usability point of views, presenting our findings in Section \ref{sec:Discussion}. We explain why our proposed model is feasible and how it helps users to configure IoT middleware platforms easily. Finally, we conclude the paper in Section \ref{sec:Conclusions}.

\section{Background and Related Work}
\label{sec:Background}

\subsection{Background}

In this section, we review a number of related work and discuss the problem domain in detail. Broadly, configuration in IoT paradigm can be categorized into two areas: \textit{sensor-level} configuration and \textit{system-level} configuration. Sensor-level configuration \cite{Z1055} focuses on changing a sensor's behaviour by configuring  its embedded software parameters such as sensing schedule, sampling rate, data communication frequency, communication patterns and protocols. In this paper, we focus on developing a system-level configuration model for IoT midddleware platforms. System-level configuration focuses on changing the behaviour of IoT middleware systems by configuring internal software components. Specifically, our proposed model identifies, composes, and configures both sensors and data processing components in order to create the data streams based on user requirements.

\begin{figure}[t!]
  \centering
  \includegraphics[scale=0.7]{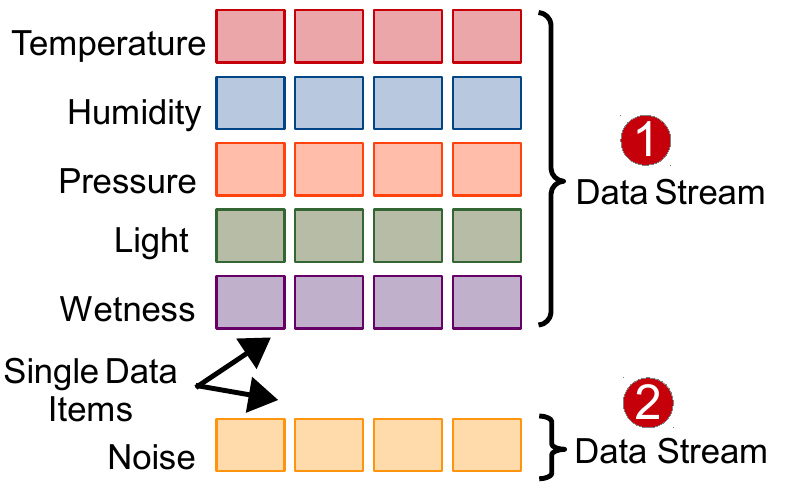}
  \caption{Data Stream}
  \label{Figure:Data_Stream}	
 \vspace{-0.23cm}	
 \end{figure}

We start by briefly introducing a reference architecture for IoT middleware. \textcolor{black}{Our reference architecture, a detailed description of which is given in \cite{ZMP004}, is presented in Figure \ref{Figure:Reference_Architecture}}. The details are presented in \cite{ZMP004}. Even though the details of this reference architecture are out of scope of this paper, we would like to briefly introduce some of the major responsibilities of an IoT middleware and its different components. The objective of an IoT middleware from users' perspective is to collect sensor data streams so they can inject them into an application that is capable of performing analysis \cite{P118}. A data stream is simply a set of data items that is captured and transferred to the users sequentially and continuously at certain intervals (e.g. every 5 seconds, every 2 hours). A sample data stream is illustrated in Figure \ref{Figure:Data_Stream}. A data stream may consist of one type of data (e.g. as illustrated in data stream 2 in Figure \ref{Figure:Data_Stream}) or multiple types of data  (e.g. as illustrated in data stream 1 in Figure \ref{Figure:Reference_Architecture}).

\textcolor{black}{The reference architecture  illustrated in Figure \ref{Figure:Reference_Architecture} consists of four layers: Data, Semantics, and Context Dissemination Layer (DSCDL), Context Processing and Reasoning Layer (CPRL), Context and Semantic Discovery Layer (CSDL), and Sensor Data Acquisition Layer (SDAL). Data, Semantics, and Context Dissemination Layer (DSCDL) is responsible for user management. The components belonging to this layer are data dispatcher, request manager, and publish/subscribe. Typically, users will not know about the technical details of the sensors or data processing components. They only know about the problem they need to solve. therefore, users need to be provided with the easy-to-use mechanisms to express their requirements in high-level without requiring technical knowledge.}

\textcolor{black}{The Processing and Reasoning Layer (CPRL)  is responsible for data processing, reasoning, fusing, knowledge generating and storing. In this layer data processing components are organized into work-flows in such a way that they collectively produce the data streams required by the consumers. Context and Semantic Discovery Layer (CSDL) is responsible for managing context and generating secondary context information from primary context information. Sensor Data Acquisition Layer (SDAL) is responsible for acquiring data.  This layer communicates with hardware and software sensors and retrieves sensor data into IoT middleware.}

\textcolor{black}{In this reference architecture, data processing components sit within the Data Fusion Operator\footnote{DFO is also called the data processing component.} (DFO) registry. Similarly, Context Provider Registry (CPR) keeps track of the data items captured by the sensors. Reasoning Engine (RE) is responsible for building the above mentioned work-flow solutions to satisfy user requests.}


\begin{table*}[b!]
\centering
\footnotesize
\caption{\textcolor{black}{Comparison of web services composition and IoT resource composition  Domains. In summary, web services selection is  based on  virtual capabilities and characteristics where IoT resources selection is based on both physical and virtual capabilities and characteristics. As a result, IoT resources are much complex elements to be selected and composed autonomously than web services.}}
\begin{tabular}{ m{0.05cm} m{8cm} m{8cm}  }

\hline  
   \multicolumn{1}{r}{}   &    
\begin{center} Web Service  Domain \cite{Zheng} \end{center} & 
\begin{center} IoT Domain \end{center} 

\\[-0.25cm] \hline \hline


\begin{sideways}Similarities\end{sideways}     
&   \begin{noindlist}
 		 \item Consuming single web service may not create significant value. Therefore, web services selection and composition is critical to generate value.
 		 \item Many alternative web services are available to use
 		 \item Can be found through directory services
 		 \item Quality of services matters \cite{Wu}.
         \item There are free as well as paid services.

    \end{noindlist}  
 &  \begin{noindlist}
  		 \item Collecting data from a single sensor may not create significant value. Therefore, sensor selection and composition is critical to generate value.
                 \item Many alternative sensors will be available to use
                 \item Middleware solutions such as OpenIoT and GSN will play a mediator roles between sensors  and sensor data consumers 
                 \item Quality of sensors (and data) matters
                 \item There will be free as well as paid sensors
 		 
     \end{noindlist}  

\\ 
\begin{sideways} Differences  \end{sideways}      
&   \begin{noindlist}
		\item Web service compose with other web services in to work-flows.
		 \item Largely guided by standards.
                 \item Largely depend on software.
                 \item Less uncertainty (unless some hardware sensors are involved. e.g, data from weather stations.)
                 \item Not tangible and more reliable.
                 \item Some web services accept data as input and produce some data based on them (e.g. data fusion).
                 \item Data send to the consumer using web services.
                 \item Comparatively, fewer web services will accessible over the Internet by 2020 \cite{Zheng}.
                 \item Typically provide more meaningful processed and refined data.

\end{noindlist}  
 &
  \begin{noindlist}
  		\item Sensors and data processing components compose together into work-flows.
 		 \item No standards (yet) \cite{Gubbi}.
         \item Largely depend on hardware, firmware, as well as software
         \item More uncertainty.
         \item Sensors are Tangible, could be mobile and less reliable. 
         \item Some sensors may accept queries/conditions/preferences as inputs and produce data based on them. Nevertheless, sensors do not accept raw data with the intention of fusing data.

		 \item Comparatively, more sensors will be accessible over the Internet by 2020 \cite{P029}.
         \item Sensors, typically provides less meaningful raw sensor data where they need to be processed by data processing components.
 \end{noindlist}  

\\[-0.25cm] \hline

\end{tabular}

\label{Tbl:Sensor Search and Webservices Domain}
\vspace{-0.33cm}
\end{table*}


The challenge of configuring an IoT middleware solution at run time can be understood by analysing an existing middleware such as Global Sensor Networks (GSN) \cite{P022}. Some of the key challenges are as follows.

\begin{itemize}
\item Users need to know the low-level details such as data types and measurement units of the sensors in order to request them manually.

\item It is extremely difficult to memorise different combinations of sensor data types that can be used to fulfil user requirements (e.g. which sensors need to be composed together to detect an event?). In particular, domain knowledge (e.g., relating to agriculture) is difficult to memorise when there are multiple ways of building a given data stream.

\item  Users need to know the availability of data processing components, their input/output data types and their capabilities  to develop a strategy. Data processing operations need to be applied on data in the correct sequence.

\item  There is no way to find out the strategies to overcome the issues when existing hardware resources (i.e. existing sensors) and software resources (i.e. data processing components) are incapable of producing the results that users required. 

\item  Further, the solutions designed by users may not be optimal (e.g. due to the variability of hardware and software costs).

\end{itemize}

An ideal IoT middleware configuration model should address all the above mentioned challenges. The proposed configuration model, CASCOM, is applicable towards several other emerging paradigms, such as sensing as a service \cite{ZMP003}. Our proposed solution combines technologies from different research areas such as IoT middleware, semantic technologies, software component composition, and context-aware computing. We discuss major related research efforts in the remainder of this section.

\subsection{Related Work}

Microsoft \textit{SensorMap} \cite{P578} (sensormap.org) is a data sharing and visualization framework. It is a peer produced sensor network that consists of sensors deployed by contributors around the world. \textit{SensorMap} mashes up sensor data on a map interface. Then, it allows to selectively query sensors and visualize data. Our approach completely automates the configuration process by eliminating the requirement of hand picking sensors. \textit{Linked Sensor Middleware} (LSM) \cite{P584} (lsm.deri.ie) is a platform that provides wrappers for real time data collection and publishing. It also provides a web interface for sensor search, linked stream data query, data annotation and visualisation. \textit{LSM} mainly focuses on linked data publishing. Sensor selection needs to be done manually in order to retrieve sensor data. \textit{Xively} (Xively.com) is a platform for Internet of Things devices. \textit{Xively} allows different data sources to be connected to it. Then, it provides functionalities such as event triggering and data filtering. It acts as a mediator between sensors and applications where users need to manually select and configure sensors. \textit{HyperCat} (hypercat.io) is an open, lightweight JSON-based hypermedia catalogue format for exposing collections of URIs. \textit{HyperCat} has proposed the notion of describing resources in a semantic way. These descriptions are designed for exposing information about IoT assets over the web. \textit{HyperCat} provides a standard mechanism for developers to publish linked-data descriptions of resources.

Context-awareness is a critical functionality that needs to be embedded into IoT middleware solutions \cite{ZMP007}. Context information (e.g. accuracy, reliability, cost) plays a significant role in selecting sensors and data processing components \cite{ZMP006}. To support this, CASCOM provides context discovery functionalities by using semantic knowledge and fusing raw sensor data. The \textit{SensorMashup} \cite{P068} platform offers a visual composer for sensor data streams. Data sources and intermediate analytical tools are described by reference to an ontology, enabling an integrated discovery mechanism for such sources. Selection of data sources and analytical tools based on user requirement need to be done manually by users. Khemakhem et al. \cite{P613} use multiple ontologies to discover and compose software components by focusing on non-functional proprieties. In web service composition domain, service composition means composing a larger service by combining many smaller services. This is the same principle used when composing a larger software component from many smaller components.

Web service (WS) composition using ontologies \cite{P601} is similar to IoT resource composition performed in CASCOM from a functional point of view but  different from an implementation and execution point of view. Web services composition domain only involves in combining multiple software components. In contrast, CASCOM needs to deal with both hardware and software components in its configuration model. We present a comparison of WS composition and IoT resource composition in Table \ref{Tbl:Sensor Search and Webservices Domain}. 

Leitner et al. \cite{Leitner} have proposed a cost-based service composition model to support service level agreements (SLA) in manufacturing domain. Similar to our approach, in SLAs, customers are allowed to express their requirements and expectations (e.g. monetary costs, time to deliver, quality). Based on the customer needs, different service providers will be used to accommodate the request order (e.g. normal shipping or express shipping). However, these composition are done based on predefined business rules. Haddad et al. \cite{Hadad} have addressed the issue of selecting and composing Web services not only according to their functional requirements but also to their transactional properties and QoS characteristics. Though QoS characteristics are important in IoT resource composition domain, transactional properties are less relevant. The reason is that IoT resource composition  does not need to support compensations or undoing transactions.  Bronsted et al. \cite{Bronsted} has mentioned that \textit{'Only 10 cases use scenario-based evaluation, which is most realistic because it involves actual use by users. So, for many of the mechanisms, there's weak empirical support for the claim that they work in realistic settings'}. This is one of the reasons we evaluate our approach using use-case and also described  the processes and techniques using real-world applications. Kritikos and Plexousakis \cite{Kritikos} have discussed quality of service and its importance towards web service discovery. They recognize QoS as a set of performance and domain-dependent attributes that has a substantial impact on WS requesters' expectations. This is also similar in IoT resource composition domain as well.

  Several projects \cite{P612} have designed and developed ontologies to describe software components. Such approaches have helped them to perform dynamic composition of software components. A process of software component matching using ontologies has been explained in \cite{P605}. In our work we employed the Software Component Ontology discussed in \cite{P612}. Semantic Sensor Ontology (SSNO) \cite{P626} also allowed us to model sensor descriptions. Noguchi et al. \cite{P615} have proposed a mechanism that generates connection between different software components in order to process sensor data and detect events. In contrast, our objective is to produce the data streams required by the users so they can be further analysed extensively using sophisticated applications.

\section{Research Challenges}
\label{sec:Problem_Analysis}

\begin{figure}[t!]
 \centering
 \includegraphics[scale=0.7]{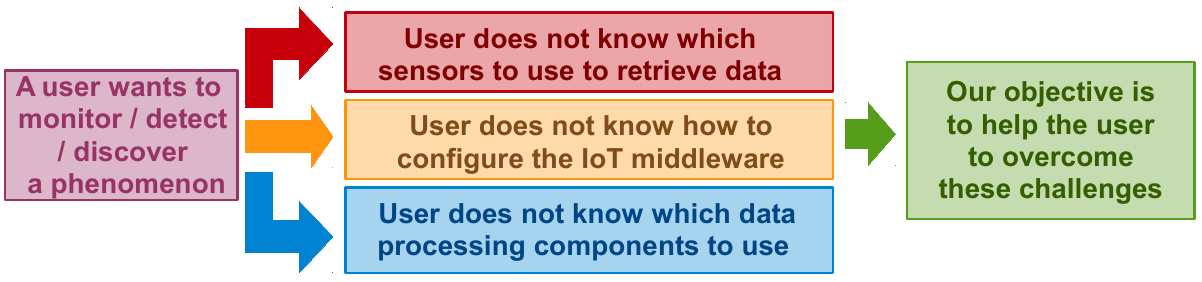}
 \caption{The Problem Definition in General}
 \label{Figure:Problem_Analysis_General}	
\end{figure}

This section describes and analyses the research challenges in detail with concrete examples and scenarios. Figure \ref{Figure:Problem_Analysis_General} illustrates the problem in general. The explanations are based on agriculture and environmental monitoring domains. The proposed solution helps data consumers to overcome the  difficulties listed in Section \ref{sec:Background}.  Our research question is \textit{`How to develop a model that allows non-IT experts to configure sensors and data processing mechanisms in an IoT middleware according to their requirements?'}.   The notations we use in this section are presented in Table \ref{Tbl:Common_Algorithmic_Notations_Resource_Composition}. Other notations we used in this paper are as follows: Wrapper ($W$) and Virtual Sensor ($VS$).

\begin{table}[t!]
\centering
\footnotesize
\renewcommand{\arraystretch}{1.40}

\caption{Common Algorithmic Notations}
\begin{tabular}{ c  m{6.5cm} }
\hline
Symbol & Definition  \\ \hline \hline

$\mathbb{S}$ & Complete set of sensors described in the data model. \\

$S_{\alpha}$ & $S$ denotes the sensor and subscript $\alpha$ denotes the types of the sensor. Examples are listed in Table \ref{Tbl:Subset_of_Sensors}.  \\ 

$\mathbb{M}_{\theta}$ & Model represent the complete ontology based semantic data model. The $\theta$ can be replaced by either $c$ as (i.e. $\mathbb{M}_{c}$) or $s$ (i.e. $\mathbb{M}_{s}$). $c$ demotes the complete model and $s$ denotes the subset of the complete model.\\

$\mathbb{T}$ & Filtered set of tasks described in the data model. \\

$T_{u}$ & Task selected by the user (or sensor data consumer) where IoT middleware needs to be configured accordingly. \\

$\Psi_{\beta}$ & SPARQL query that selects different properties from the data model. $\beta$ can be replaced by $t$ tasks, $a$ answers, $q$ questions, and $d$ data streams.   \\

$\mathbb{Q}$ & Filtered  set of questions described in the data model (List of questions).  \\
$Q_{u}$ & Single question selected by the user to answer.  \\

$\mathbb{A}$ & Filtered  set of answers described in the data model (List of answers).  \\
$A_{u}$ & Single answer selected by the user.  \\

$C_{\gamma}({\Delta}):z $ & $C$ denotes the data processing components where $\gamma$ is used an identifier to distinguish each different component. Arguments/ parameters accept by each components are depicted   by $\Delta$ as set. $\Delta$ may accept one or more inputs as denoted by `$\lambda_{\#}$'. The symbol \# denotes the number of the input parameter. The type of each argument is depicted by letters such as $x, y$ (i.e. $\Delta=\{\lambda_{1} x, \lambda_{2} y\}$). The return value is depicted by letter after `$:$' symbol. Examples are listed in Table \ref{Tbl:Subset_of_Components}.\\

$\mathbb{H}$ & Filtered  set of solutions composed by CASCOM which are capable to producing data streams required by the user.   \\

$H_{u}$ & A single solution composed by CASCOM which are capable to producing data streams required by the user. Solutions is a composition of sensors and data processing components formulated into a certain order.  \\

$\mathbb{D}$ & Filtered  set of different data-streams that can fulfil user requirements. Data stream is a continuous flow of data which encompasses several data items.  \\

$D$ &  A single data stream is composed with number of different data items.  \\

$\varPi_{i}$ &  This denotes the $i^{th}$ data item of a given data stream. \\

$\mathbb{R}$ & Recommendation list that contains information about sensors and data processing components that are not available. Acquire such resources will  help to facilitates user requirement in the future.   \\

$\mathbb{P}$ & List of all the data items that are available to be captured by the IoT middleware either directly through active wrappers or by combining / composing such data iteams with data processing components. Depending in the complexity of retrieving and generating data items, we categorize them into number of categories. \\

$P$ &  A single data item that is available to be captured from active wrapper.\\

$\mathbb{M}$ &  Additional list context information that can be discovered by the users if needed. \\

$\sigma$ &  Matrix that stores the information about input / outputs of data processing components and data items in $\mathbb{P}$. \\

\hline
\end{tabular}
\label{Tbl:Common_Algorithmic_Notations_Resource_Composition}
\end{table}

 \begin{table}[t!]
 \centering
 \footnotesize
 \renewcommand{\arraystretch}{1.25}
 \parbox{.45\linewidth}{
 \centering
 \begin{tabular}{ c  l }
 \hline
 Sensor  & Explanation  \\ \hline \hline
 
 $S_{AT}$ & Air Temperature  \\ 
 $S_{AH}$ & Air Humidity   \\ 
 $S_{LW}$ & Leaf Wetness   \\ 
 $S_{CM}$ & Carbon Monoxide   \\ 
 $S_{CD}$ & Carbon Dioxide   \\ 
 $S_{MO}$ & Molecular Oxygen   \\ 
 $S_{ME}$ & Methane   \\ 
 $S_{ND}$ & Nitrogen Dioxide   \\ 
 \hline
 \end{tabular}
 \caption{Subset of Sensors}
 \label{Tbl:Subset_of_Sensors}
 }
 \hfill
 \parbox{.45\linewidth}{
 \centering
 \begin{tabular}{ c  l }
 \hline
 Sensor  & Explanation  \\ \hline \hline
 
$C_{1}({\Delta}):z $  & airStressDetector  \\ 
$C_{2}({\Delta}):z $& phytophtoraMonitor   \\ 
$C_{3}({\Delta}):z $ & pollutionDetector   \\ 
$C_{4}({\Delta}):z $ & airQualityMonitor   \\ 
 \hline
 \end{tabular}
 \caption{Subset of DPCs}
  \label{Tbl:Subset_of_Components}
 }
 \end{table}

\begin{figure}[b!]
 \centering
 \includegraphics[scale=.7]{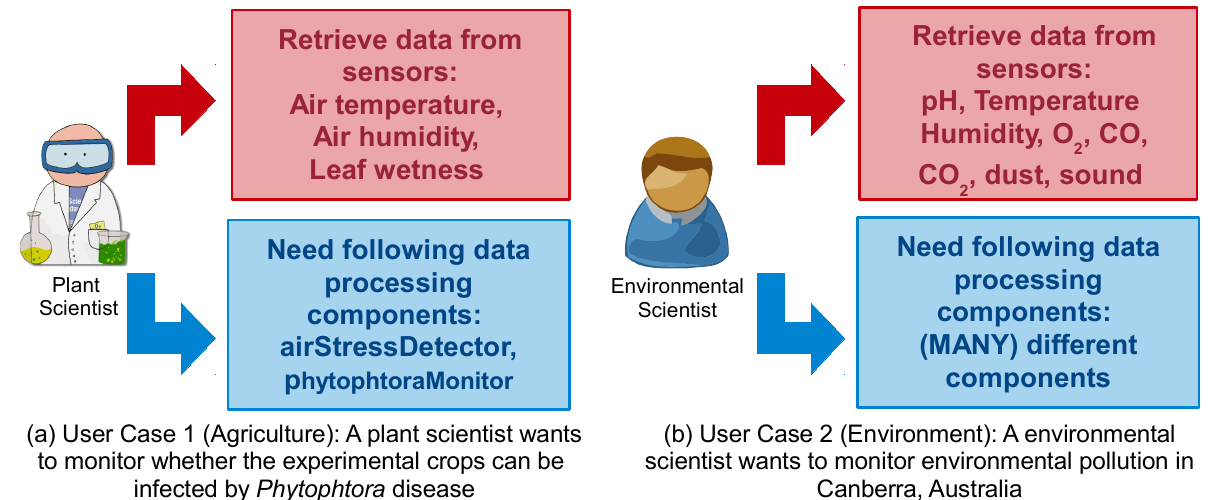}
 \caption{Use cases that illustrates the need of CASCOM.}
 \label{Figure:Problem_Analysis}	
\end{figure}

Figure \ref{Figure:Problem_Analysis} illustrates two scenarios from two different domains. Each of them has different  consumer requirements that lead to two different execution flows. We selected these two scenarios due to the fact that, together, they allow us to showcase the full capabilities of  Context-Aware Sensor Configuration Model (CASCOM). In use case 1, a plant scientist wants to monitor whether the experimental crops can be infected by \textit{Phytophtora} \cite{P452} disease or not. \textit{Phytophtora} is a fungal disease which can enter a field through a variety of sources. The development and associated attack of the crop depends strongly on the climatological conditions within the field. Humidity plays a major role in the development of \textit{Phytophtora}. Both temperature and whether or not the leaves are wet are also important indicators to monitor \textit{Phytophtora}. The following facts explain \textit{Phytophtora} monitoring (simplified for demonstration purposes). It is important to highlight that rule-based reasoning\footnote{\textcolor{black}{we employed rules based reasoning in this disuccions}} does not intended to replace rule engines \cite{P598}. The objective here is to create the data items that are required by the application.

\begin{itemize}
 \item IF \textbf{Air Temperature}  $\textless$ $\alpha$ AND \textbf{Air Humidity} $\textless$ $\beta$ THEN \textbf{Air Stress} level = low ELSE \textbf{Air Stress} level = high

 \item IF \textbf{Air Stress} = high  AND \textbf{Leaf Wetness} $\textgreater$ $\delta$ THEN \textit{Phytophtora Disease} = Can-be-infected ELSE = Cannot-be-infected

\end{itemize}

One of the responsibilities of an IoT middleware is to combine different sensors and data processing components autonomously and produce a data stream as illustrated in Figure \ref{Figure:Data_Stream}. Mostly, our focus is on data streams that consists of multiple data types. A data consumer can feed the data stream into an application for further complex processing such as visualization and modelling that allows the data consumers to achieve their objectives. The main challenge is that the plant scientist may not know (or remember) the domain knowledge listed as rules above. Further, we should not expect a plant scientist to write XML or Java code as part of the configuration. An ideal IoT middleware should help the scientist (non-IT expert) to overcome these challenges by providing tools that are easy to use. The scientist should be able to configure the middleware according to the problems/tasks at hand with minimum effort. Additionally, advanced customization will be useful to optimize the configuration process. Comparatively, use case 1 is less complex as there is only one way to monitor the disease (above rules). For example, the sensor types and data processing components need to be used are straight forward. 

\begin{itemize}

 \item \textbf{Use case (1) Solution:} $   \left (( S_{AT}, S_{AH} )  \Rightarrow  C_{1}, S_{LW} \right )\Rightarrow C_{2}  $

\end{itemize}

\begin{figure}[h!]
 \centering
 \includegraphics[scale=0.85]{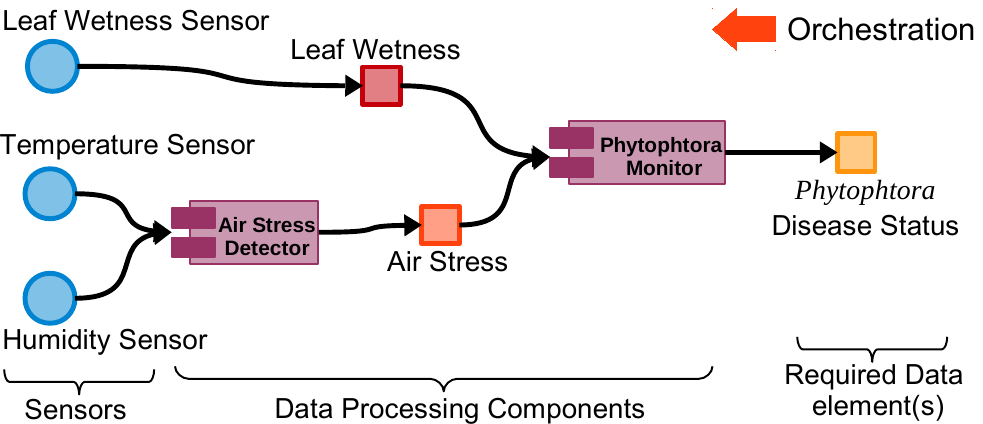}
 \caption{Resource Composition in IoT}
 \label{Figure:IoT_Resource_Composition}	
\end{figure}

As symbolized in the above statement, the IoT resource may need to be composed as follows. First, air temperature ($S_{AT}$) and air humidity ($S_{AH}$) need to be fed into \textit{airStressDetector} component ($C_{1}$). Then, it produces the \textit{airStress} as the outcome. Then leaf wetness ($S_{LW}$) and airStress need to be fed into \textit{phytophtoraMonitor} component ($C_{2}$). It produces the \textit{phytophtoraDisease} status. The IoT resource composition is illustrated in Figure \ref{Figure:IoT_Resource_Composition}

Configuration becomes a complex task in the use case 2. In this scenario, an environmental scientist wants to measure the environmental pollution in Canberra, Australia. In comparison to the use case 1, there are many different ways to measure and visualize pollution. Different sensors and data processing components can be combined together to fulfil the requirements of data consumers as listed below. Even the same {Data Processing Components} (DPC) may accept different combination of input in order to perform the same task. \textcolor{black}{DPC is a black box that accept certain types of inputs and produces certain types of outputs. The reasoning happen within a given DPC could be varies from, rules based reasoning, statistical reasoning, logical inferencing machine learning, probabilistic reasoning and so on. As an example, we used a rule based DPC in the paper discussion.}


\begin{itemize}

  \item \textbf{Use case (2) Solution 1:} $   \left ( S_{CM}, S_{CD} , S_{MO} , S_{ME} , S_{ND}\right ) \Rightarrow  C_{4}  $

  \item \textbf{Use case (2) Solution 2:} $   \left ( S_{CD}, S_{ND}  \right ) \Rightarrow  C_{3}  $

  \item \textbf{Use case (2) Solution 3:} $  \left ( S_{AT}, S_{CD}, S_{ME} \right ) \Rightarrow  C_{4}  $

\end{itemize}

\begin{figure}[b!]
 \centering
 \includegraphics[scale=.72]{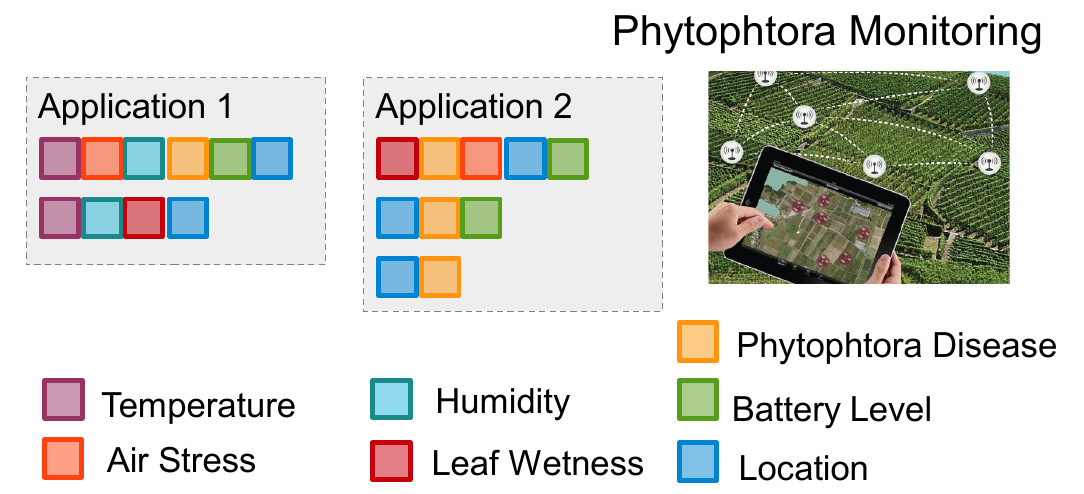}
 \caption{Each application may accept different data streams and provide  outputs at different detail levels}
 \label{Figure:Different_Datastreams}	
\end{figure}

In such circumstances, it is important to consider context information (e.g. accuracy, reliability) and cost of data acquisition (e.g. data communication time and computation time). The availability of more than one option allows a data consumers to make the final decision on which solution to be used depending on the cost and context factors. Both hardware and software costs need to be considered. Additionally, data consumers may need to discover additional context information \cite{ZMP007}. Depending on the requirements of the data consumers and  application requirements, the required output data stream may vary. Sample data streams, in relation to use case 1, are listed below.

\begin{itemize}

  \item \textbf{Output 1:} airTemperature [double], airHumidity [double], airStress [string], \\leafWetness [double], PhytophtoraDisease [boolean]
  
  \item \textbf{Output 2:} PhytophtoraDisease [boolean], location [string], batteryLevel[double]

\end{itemize}

Previously, we explained that the objective of IoT middleware is to produce data streams so the users can inject them into applications. We assume that these applications will accept multiple different data streams as illustrated in Figure \ref{Figure:Different_Datastreams}. The ideology is that when these applications are provided with more data items, they will perform better or provide additional featured / results. However, each application will have a minimum number of data items that it would accept in order to perform the primary task it promises to deliver. For example, if the application 2 (in Figure \ref{Figure:Different_Datastreams}) is provided with \textit{PhytophtoraDisease} and \textit{location} data, it will simply  mark the areas which are under risk of getting infected by  \textit{PhytophtoraDisease}. In contrast, if the application 2 is provided with  more information such as \textit{batteryLevel}, \textit{leafWetness}, and \textit{airStress}, it will produces more detailed and comprehensive visualization that may include risk level  with certain confidence. Raw data values of \textit{leafWetness}, and \textit{airStress} may help the application 2 to perform these additional calculations and predictions.

In summary, we assume each application would perform one or more tasks (e.g. \textit{PhytophtoraDisease}  monitoring). Each application would accept one more  different data streams. In such circumstances, each data stream may consist of different types of data items. Additionally, different developers (or companies) or the same developer may develop multiple applications that perform the same tasks. Similarly, we also assume that there would multiple data processing components that would perform the same data fusion operations though their context information may varied. Due to the  large number of possibilities, IoT middleware platforms  require an automated process to optimally serve the user requests.

\section{Architectural Design}
\label{sec:Architectural_Design}
 
\begin{figure}[b]
 \centering
 \includegraphics[scale=.72]{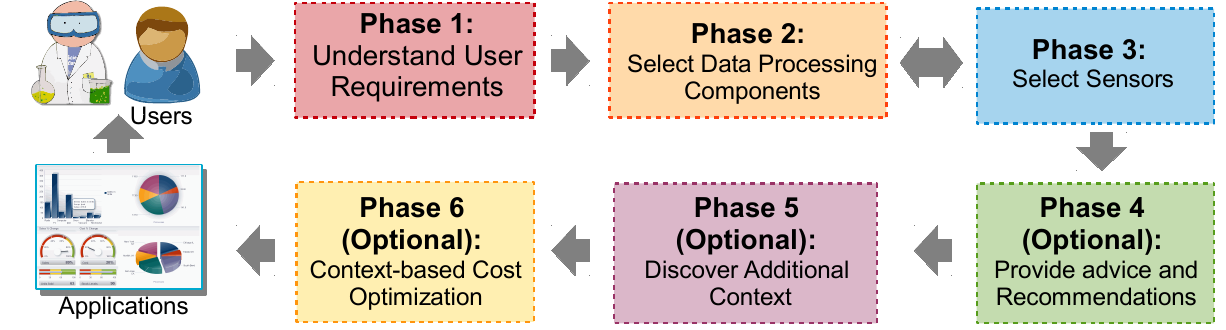}
 \caption{The Context-Aware Sensor Configuration Model (CASCoM)}
 \label{Figure:The_CASCOM_Model}	
\end{figure}

Based on the challenges we identified in Section \ref{sec:Problem_Analysis}, we designed a model, which is supported by a tool, to overcome the difficulties. Context-Aware Sensor Configuration Model (CASCOM) simplifies the IoT middleware configuration process significantly. Our proposed model allows non-technical personnel to configure IoT middleware effortlessly. All the technical configurations are handled internally behind the scenes without the users' involvement. Additionally, we offer several advanced features that allow optimization and customizations. As depicted in Figure \ref{Figure:The_CASCOM_Model}, CASCOM consists of six phases. Some phases may or may not be visible to the users. Phases are different from the steps needed to be followed in the CASCOM Tool.

\textbf{CASCOM Execution Flow:} In phase 1, data consumers (users) interact with a graphical user interface that is based on a \textit{question-answer (QA)} approach, to specify their requirements. Users can answer as many questions as possible. CASCOM searches and filters the tasks that the user may want to perform. From the filtered list, users can select the desired task. The details of the QA approach are presented later in this section. In phase 2, CASCOM searches for different programming components that allow to generate the data stream required. In phase 3, CASCOM tries to find the sensors that can be used to produce the inputs required by the selected data processing components. If CASCOM fails to produce the data streams required by the users due to insufficient resources (i.e. unavailability of the sensors), it will provide advice and recommendations on future sensor deployments in phase 4. Phase 5 allows the users to capture additional context information. The additional context information that can be derived using available resources and knowledge are listed to be selected.  In phase 6, users are provided with one or more solutions\footnote{A solution is a combination of sensors and data processing components that can be composed together in order to satisfy the user requirements.}. CASCOM calculates the costs for each solution. By default, CASCOM will select the solution with lowest cost. However, users can select the cost models (discussed later in this section) as they required. Finally, CASCOM generates all the configuration files and program codes which the actual IoT middleware may requires\cite{ZMC007}. Data starts streaming soon after. 

 \begin{algorithm}[b!]
 
 \algsetup{linenosize=}
 \caption{Question-Answer based Task Filtering}
 \label{Alg:Question_Answer_based_Task_Filtering}
 \begin{algorithmic}[1]

 \REQUIRE ($\mathbb{M}$).  
 \STATE \textbf{Output:}  $T_{u}$
 \STATE $\mathbb{M} \leftarrow \textrm{Load data into model}$
 \WHILE{$T_{u}\textrm{!= NULL}$}
	 \STATE $\mathbb{Q} \leftarrow executeQuery(\Psi_{q},\mathbb{M})$
	 \STATE $Q_{u} \leftarrow \textrm{Ask user to select a question from } \mathbb{Q}$
	 \STATE $\textrm{Add } Q_{u} \textrm{ to }  \Psi $
	 \STATE $\mathbb{A} \leftarrow  executeQuery(\Psi_{a},\mathbb{M})        $
	 \STATE $A_{u} \leftarrow \textrm{Ask user to select a answer from } \mathbb{A}$
	 \STATE $\textrm{Add } A_{u} \textrm{ to }  \Psi $
	 \STATE $\mathbb{T} \leftarrow  executeQuery(\Psi_{t},\mathbb{M}) $ 
	 \STATE $T_{u} \leftarrow \textrm{Ask user to select a task from } \mathbb{T}$
	 \IF{$T_{u}\textrm{!= NULL}$} 
	 \RETURN $T_{u}$  
	 \ENDIF
 \ENDWHILE
 \end{algorithmic}
 \end{algorithm}

\begin{figure}[t!]
 \centering
 \includegraphics[scale=1.1]{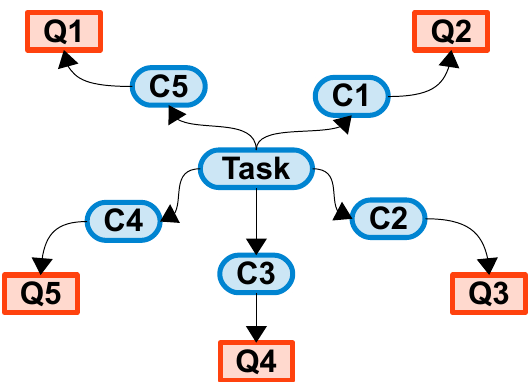}
 \caption[Question-Answer Filtering supported by QA-TDO ]{A part of QA-TDO shows how we developed the QA model. It is important to note the pattern (i.e. \textit{Task} $\rightarrow$ \textit{Concept} $\rightarrow$ \textit{Question}).}
 \label{Figure:QA}	
\vspace{-0.30cm}	
\end{figure}

\textbf{Phase 1: Understand User Requirements:} The objective of this phase is to help  data consumers to search for a task (e.g. \textit{PhytophtoraDisease}  monitoring) that they need to perform easily from a large number of possibilities. For example, data consumers are allowed to narrow down the possibilities by mentioning facts such as domain (e.g, agriculture), and type of the task (e.g. event, visualization). In order to increase the usability, CASCOM retrieves the facts from the data consumers through a QA model (Sample questions: Do you want to visualize data?, Do you want to detect an event?, Do you want to monitor a disease infection? What is the domain your task is related to?). When a user answer a question, the remaining questions will be dynamically selected based on the previous answer. An extract of the proposed \textit{Question and Answer oriented Task Description Ontology} (QA+TDO) is presented in Figure \ref{Figure:QA}. In QA+TDO, tasks can be explained by any concept as depicted in $C1$, $C2$, etc. in Figure \ref{Figure:QA}. Each concept should have a \textit{`hasQuestion'} property which links to a question (i.e. $Q1$, $Q2$ and so on). It is expected that new questions will be added to the QA+TDO over time by different domain experts and contributors as part of the knowledge modelling process so the non-technical users can take advantage of them. In QA+TDO, $C$  are answers to the questions. (e.g, If $Q1$= What is the domain your task is related to?, then $C5$ is \textit{`domain'} and an individual of $C5$ can be \textit{`agriculture'}.).  This process is presented in algorithmic perspective in Algorithm \ref{Alg:Question_Answer_based_Task_Filtering}, which takes the data model as the input and outputs the user preferred task. The design philosophy of this algorithm is that it repeatedly allow the user to select questions and answer them. As a result, the algorithm will generate a SPAQRL statement and update it every time when a user selects and answers a question. Every time a user answers a question, the number of possible options offered to the users will get reduced as the new Q\&A will always add more constraints to the query.

\textcolor{black}{Let us briefly explain the algorithm \ref{Alg:Question_Answer_based_Task_Filtering}. CASCOM first allows users to select a question as demoted in $Q_{u}$. Next CASCOM uses $Q_{u}$ to query its knowledge-base. The results are denoted as $\mathbb{A} $. Next, both question and answers is amended to a single query $\Psi $. $\Psi $ is used to query the knowledge-base and resulted  tasks are denoted by $\mathbb{T}$. This process repeats until users find the tasks they are looking for $T_{u}$.}

\begin{figure*}[t!]
 \centering
 \vspace{-0.23cm}
 \includegraphics[scale=1.15]{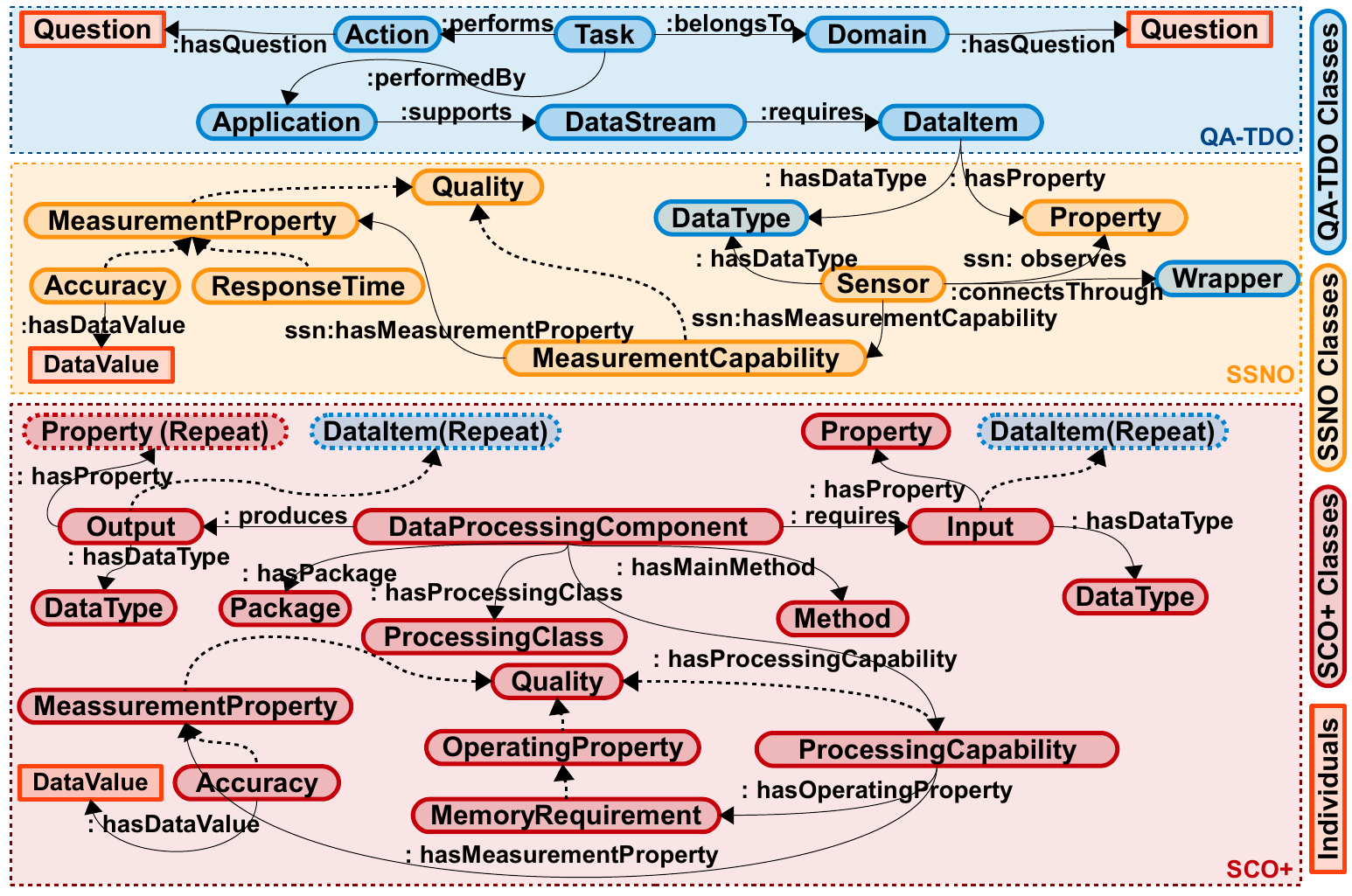}
 \caption[Proposed Data Model for CASCOM]{Extracts of different ontological data models used in CASCOM: QA-TDO, SCO \cite{P612}, and SSN ontology \cite{P626}. The colour coding refers to different prefixes. Prefixes in abbreviated Internationalized Resource Identifiers (IRIs).}
 \label{Figure:Ontology_Model}	

\end{figure*}

\begin{algorithm}[t!]
 
 \algsetup{linenosize=}
 \caption{Resources Composition and Recommendation}
 \label{Alg:On_Demand_IoT_Resources_Composition}
 \begin{algorithmic}[1]

 \REQUIRE ($\mathbb{M}$), ($T_{u}$)
 \STATE \textbf{Output:}  $\mathbb{H}, \mathbb{R}$
 \STATE $\mathbb{M} \leftarrow \textrm{Load data into model}$
 \STATE $\mathbb{D}  \leftarrow executeQuery(\Psi_{d}, T_{u}, \mathbb{M})$
 \STATE$//D = \{\varPi_{1},\varPi_{2},\varPi_{3}...\varPi_{n} \}$
 \FORALL{ $D \in \mathbb{D}$} 
 	\FORALL{ $\varPi_{i} \in D$}  
    	 \IF{$\varPi_{i} == \textrm{output of a sensor } S_{\alpha} \textrm{ in the set } \mathbb{S}  $} 
    	 	\STATE $\textrm{add } S_{\alpha} \textrm{ to } H_{u}$  
    	 \ELSE \IF {$\varPi_{i} == \textrm{output of a component } C_{\gamma} \textrm{ in the set } \mathbb{C}  $} 
    	    \STATE $\textrm{add } C_{\gamma} \textrm{ to } H_{u}$ 
    	    \STATE $composeFurther(C_{\gamma}, H_{u}) $
    	        \ELSE
    	          	  \STATE $\textrm{add } \lambda_{\#} \textrm{ to } \mathbb{R}$      	      	    
    	 \ENDIF
    	 \ENDIF
 	\ENDFOR 
 \ENDFOR
 \STATE 
 \STATE $ \textrm{\textbf{Function} } composeFurther (C_{\gamma}, H_{u})$
  \FORALL{ $\lambda_{\#} \in \Delta \textrm{ \textbf{of} } C_{\gamma}$} 
  	 \IF{$\lambda_{\#} == \textrm{output of a sensor } S_{\alpha} \textrm{ in the set } \mathbb{S}  $} 
      	 	\STATE $\textrm{add } S_{\alpha} \textrm{ to } H_{u}$ 
      	 	\STATE $break$
      		 \ELSE \IF {$\varPi_{i} == \textrm{output of a component } C_{\gamma} \textrm{ in the set } \mathbb{C}  $} 
      	   		\STATE $\textrm{add } C_{\gamma} \textrm{ to } H_{u}$ 
      	    	\STATE $composeFurther( C_{\gamma}, H_{u}) $
      	    \ELSE
      	    	\STATE $\textrm{add } \lambda_{\#} \textrm{ to } \mathbb{R}$      	      	    
      	 \ENDIF
      	 \ENDIF
   \ENDFOR
 \end{algorithmic}
 \end{algorithm}

\textbf{Phase 2 and 3: Select Sensors and Data Processing Components:} CASCOM requires all the information related to sensors and data processing components to be stored in a repository. We extended the Software Component Ontology \cite{P612} (SCO) as presented in Figure \ref{Figure:Ontology_Model} in order to model information about data processing components. Further, we modelled sensor descriptions using semantic Sensor Ontology (SSN) \cite{P626}. In this phase, the software components are selected in such a way that they can together produce the data stream required to perform the task selected in phase 1. For example, in order to monitor \textit{PhytophtoraDisease}, first CASCOM searches for a software component that can be used to produce the required data. It first finds  \textit{PhytophtoraDisease Detector}.  The inputs it requires are \textit{air stress} and \textit{leaf wetness}. Phase 3 selects the sensors that produce the output that matches the inputs of the selected component. \textit{Leaf wetness} can be measured directly using hardware sensors. However, \textit{air stress} cannot be detected using any physical sensor. This requires CASCOM to execute phase 2 again in order to find a software component that produces \textit{air stress}. Then CASCOM finds  \textit{Air Stress Detector} which takes \textit{air temperature} and \textit{air humidity} as inputs and produces \textit{air stress} as the output. Further, \textit{air temperature} and \textit{air humidity} can be sensed directly through hardware sensors. The IoT middleware configuration process will be completed once the required sensors and data processing components are identified. The remaining phases are optional. 

\begin{figure}[t!]
 \centering
 \includegraphics[scale=1.1]{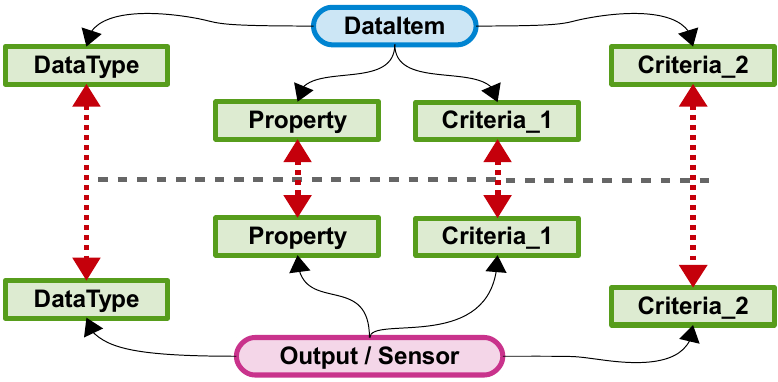}
 \caption[Validation in Resource Composition]{IoT Resource compositions need to be validated before presented to the users. Semantic meanings as well as syntactic definition (e.g. programming level data types) need to matched and compatible}
 \label{Figure:Valiication}	
\vspace{-0.12cm}	
\end{figure}

CASCOM performs validation as illustrated in Figure \ref{Figure:Valiication}. During the sensors and data processing components composition process, different criteria are evaluated  (e.g. data types: int, boolean / measurement units: Celsius, Fahrenheit) in order to verify whether the inputs and outputs are compatible. The above mentioned procedures are presented in algorithmic perspective in  Algorithm \ref{Alg:On_Demand_IoT_Resources_Composition}.

\textcolor{black}{This algorithms takes the data model (demoted by $\mathbb{M}$) and the data stream elements of the user preferred task as the inputs (demoted by $T_{u}$). First, it finds out what are the output data stream required by the user preferred task (denoted by $\mathbb{D} $). Then, it attempts to generate that data stream by composing sensors and data processing components (from line 5-18). The design philosophy behind the search and composition is that priority is given to prepare the output data stream using direct sensor outputs (denoted by $S_{\alpha}$). If this is not possible (e.g. when a certain data element cannot be directly sensed), the algorithm will search for a data processing component which may be able to produce the required output (denoted by $C_{\gamma}$). If it succeeds, then the inputs of the selected data processing component will be searched (using the $composeFurther(C_{\gamma}, H_{u}) $). As illustrated in Figure \ref{Figure:IoT_Resource_Composition}, this process will continue until the algorithms finds ways to produce the elements in the required output data stream.}

\textbf{Phase 4 (Optional): Provide Advice and Recommendations:} Through comparing SSN ontology and SCO, this phase identities the resource insufficiencies and provides advice to the data consumers regarding future sensor deployments and software component acquisition. This phase provides alternative advice if there are multiple ways to address the insufficiencies (e.g. use case 2). As presented in Algorithm \ref{Alg:On_Demand_IoT_Resources_Composition}, resource insufficiencies are also detected and identified during the resource composition process. A list of resource insufficiencies is prepared and returned as $\mathbb{R}$.

\textcolor{black}{Lets consider use case 2. Its objective is to determine environmental pollution in a city. As presented in Section \ref{sec:Problem_Analysis}, there are three different solutions that that can be used to achieve this objective. Assume, in our IoT system, we only have  access to sensors $    S_{CD} $ and $ S_{ME} $ . However, those two sensors are not capable of producing data that is required by any of the  exiting DPCs, namely $C_{3}$ and $C_{4}$. Therefore, this phase of our model recommends users to deploy either sensors $ S_{ND} $ or $ S_{AT} $. Such deployments will fulfil the data requirements of above components. }

 \begin{algorithm}[t!]
  
  \algsetup{linenosize=}
  \caption{Context Discovery}
  \label{Alg:Context_Discovery}
  \begin{algorithmic}[1]
 
  \REQUIRE (List of Active Wrappers).  
  \STATE \textbf{Output:}  $\mathbb{M}$
   \STATE $\mathbb{P} = \{\mathbb{P}^{0}, \mathbb{P}^{1}, \mathbb{P}^{2}... \mathbb{P}^{n}\}$
  \STATE $\mathbb{P}^{0} \leftarrow \textrm{List data items available through active wrappers}$
   \STATE $ \textrm{add } \mathbb{P}^{0} \textrm{ to } \mathbb{M} $
   \FORALL{ $\mathbb{P}^{i} \in \mathbb{P}^{n}$} 
 	 \FORALL{ $C \in \mathbb{C}$} 
 	 	\FORALL{ $ \lambda_{\#} \in C$}
 	 	 \IF{$\lambda_{\#} == \textrm{any } P \textrm{ in the set } \mathbb{P}^{i}  $} 
 	 		 \STATE $\textrm{add } \checkmark \textrm{ to } \lambda_{\#} \textrm{ of } C \textrm{ in } \sigma$  
 	 	 \ENDIF
 	 	 
 	 	\ENDFOR
   \ENDFOR
   \FORALL{ $C \in \mathbb{C}$} 
   	 \IF{$ \textrm{All }\lambda_{\#}  \textrm{ of } C == \checkmark $}
   	 	\STATE $\textrm{add output of } C \textrm{ to } \mathbb{P}^{i+1}$  
   	 	\STATE $\textrm{add } \checkmark \textrm{ to } \lambda_{\#} \textrm{ of } C \textrm{ in } \sigma$
   	 \ENDIF 
   \ENDFOR
  \ENDFOR 
    \FORALL{ $C \in \mathbb{C}$} 
    	 \IF{$ \textrm{All }\lambda_{\#}  \textrm{ of } C == \checkmark $}
    	 	\STATE $\textrm{add output of } C \textrm{ to } \mathbb{M}$  
    	 \ENDIF 
    \ENDFOR

  \end{algorithmic}
  \end{algorithm}
\textbf{Phase 5 (Optional): Additional Context Discovery:} With the help of knowledge modelled in ontologies, this phase discovers context information that can be derived by using sensor data. Additional context information such as sensor location and sensor battery life may be required by applications in order to perform complex tasks such as geographical visualization and developing energy-aware sensing schedules. Therefore, discovering additional context is important. Each application may have a compulsory set of inputs that it needs to perform the primary task, though they may accept additional context information in order to provide enhanced results. 

\begin{figure}[h!]
 \centering
 \vspace{-0.12cm}
 \includegraphics[scale=.70]{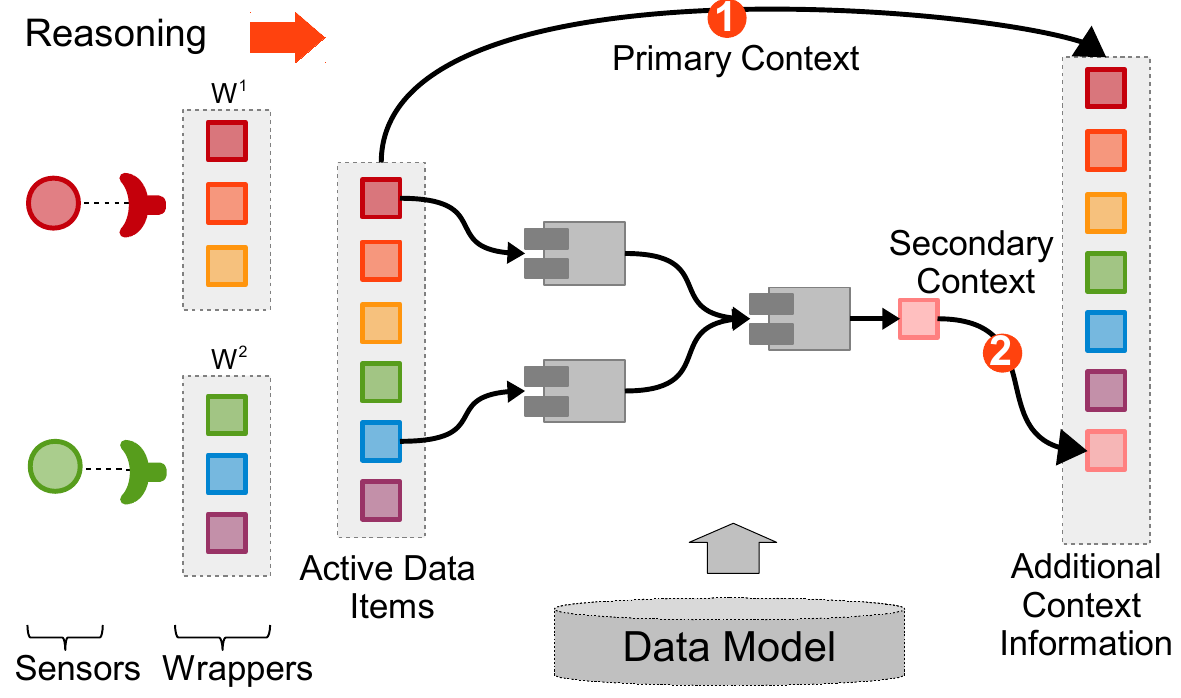}
 \caption{Primary and Secondary Context Discovery}
 \label{Figure:Context_Discovery}	
\vspace{-0.63cm}	
\end{figure}

\textcolor{black}{First, all the data items directly retrieved through sensors (we call them parameters and denoted by $\mathbb{P}$) are added to a list of context information  denoted by $\mathbb{M}$. Such pieces of context information are referred to as primary context \cite{ZMP007}. Each  wrapper has set of data items (i.e. parameters) it can produce. Set of parameter produce by each active wrapper is noted by $ \{\mathbb{P}^{0}, \mathbb{P}^{1}, \mathbb{P}^{2}... \mathbb{P}^{n}\}$.   Next, these primary context parameters are composed with all existing DPCs (denoted by $\mathbb{C}$) to check whether secondary context \cite{ZMP007} can be produced. If possible, such secondary context parameters are also added to the list of context information  denoted $\mathbb{M}$ as well. The context discovery procedures are presented in algorithmic perspective in self-explanatory Algorithm \ref{Alg:Context_Discovery}. Further, We can explain the context discovery procedure using the Figure \ref{Figure:Context_Discovery}. }  This algorithm works independently from  users' preferences. Further, it can also be preprocessed. The design philosophy behind this algorithm is that it attempts to identify all possible secondary context information that can be generated by combining all the possible outputs of sensors as well as data processing components. This algorithm is only required to run when a new sensor or data processing component appears or when existing resource disappears.

\textbf{Phase 6 (Optional): Context-based Cost Calculation:} 
In CASCOM, the main objective is to identify the required IoT resources at a conceptual-level. In the first 5 phases, we achieve this main objective. In phase 6, we focus on identifying actual IoT resources. It is important to note that there can be multiple DPCs that can perform similar tasks. Further, there are large numbers of sensors available with overlapping and sometimes redundant functionality. In such situation, data consumers may want to decide the exact criteria that IoT resources selection process should consider.

CASCOM performs ontological reasoning to find out all possible solutions. Each solution may combine different sensors and data processing components where their costs may different. For example, different types of sensors can be used to monitor environmental pollution as illustrated in Figure \ref{Figure:Problem_Analysis}. Cost does not always refer to financial terms (e.g, sensors: energy, bandwidth, latency; data processing: memory requirement, processing time). By default, all the context parameters are treated equally. However, users can define their priorities for each context property in comparative fashion \cite{ZMP006}. If the  users want more \textit{reliable} sensors, the \textit{reliability} can be defined with more priority, but it may increase the cost.

 \begin{figure*}[t!]
  \centering
  \includegraphics[scale=1.2]{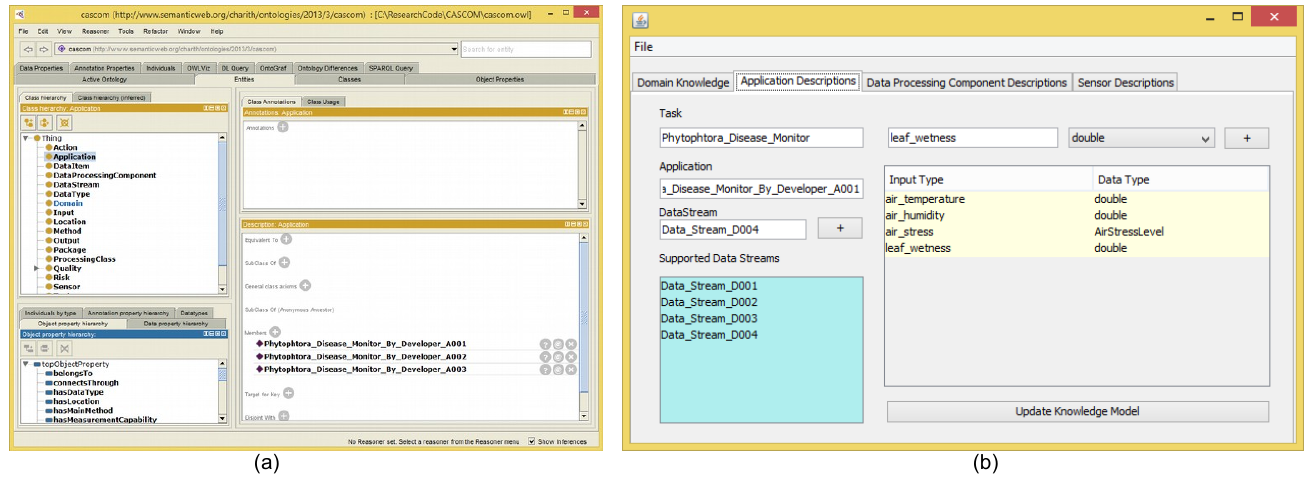}
  \vspace{-0.23cm}	
  \caption{Semantic Data Modelling Tools: (a) \textit{Protege}  and (b) proposed IoT resource description tool}
  \label{Figure:Support_Tool}	
 \end{figure*}

 \begin{figure*}[b!]
  \centering
  \includegraphics[scale=1.4]{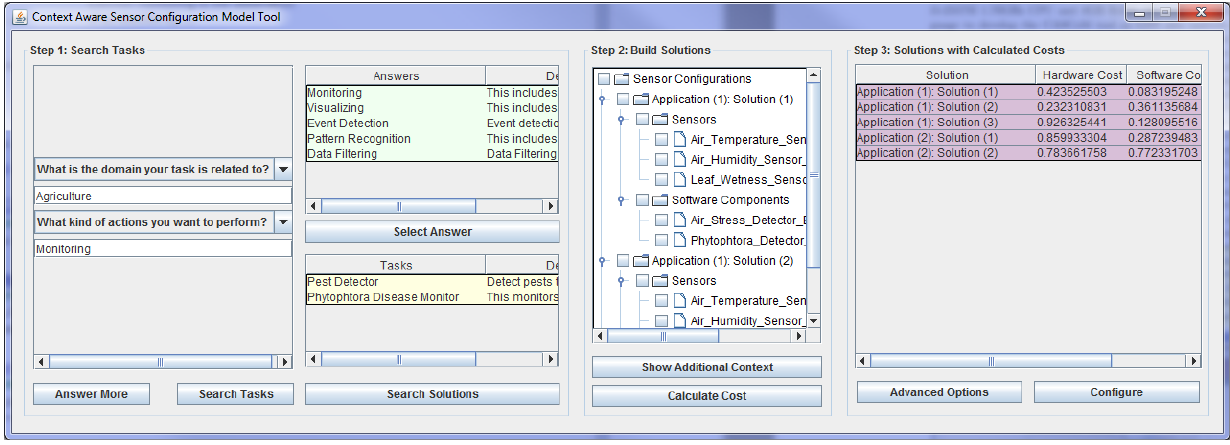}
  \vspace{-0.23cm}	
  \caption{User interface of the software tool that supports CASCOM}
  \label{Figure:User Interface1}	
 \end{figure*}

 \section{Description Generation Tool}
 \label{sec:Support_Tools}
 
 In our proposed model, the description of IoT resources and related knowledge play a significant role. Today, even though there are sophisticated tools that can be used to develop \textit{ontologies} and model \textit{instances} such as  \textit{Protege} \cite{Protege}, they are very complex to use. The learning curve of these tools are  significant. The user interface of \textit{Protege} tool, with CASCOM data model opened, is presented in Figure \ref{Figure:Support_Tool}(a). As it is clearly visible, the \textit{Protege} user interface looks very complex to someone who has never used it before and hard to understand where to even begin.
 
  In CASCOM, we expect data processing components, sensors, and domain knowledge to be collectively described and modelled by developers, domain experts, and non-technical personnel (e.g. capabilities, inputs, output, etc.). However, not even all developers are familiar with semantic modelling tools such as \textit{Protege}. Therefore, we built a very simple form-based tool where anyone can learn and use it with very limited effort. They can fill the form and the tool will model the according to CASCOM ontology behind the scenes. More importantly, this tool can be used to add IoT resource descriptions to existing knowledge models. Our form-based knowledge modelling tool in presented in Figure \ref{Figure:Support_Tool}(b). This tool consists of number to separate tabs where each tab allows users to model certain type of knowledge (e.g. describe a data processing component, describe sensors, add domain knowledge, etc.).

\begin{figure*}[b!]

        \centering
        \begin{subfigure}[b]{165pt}
                \centering
                \includegraphics[scale=0.34]{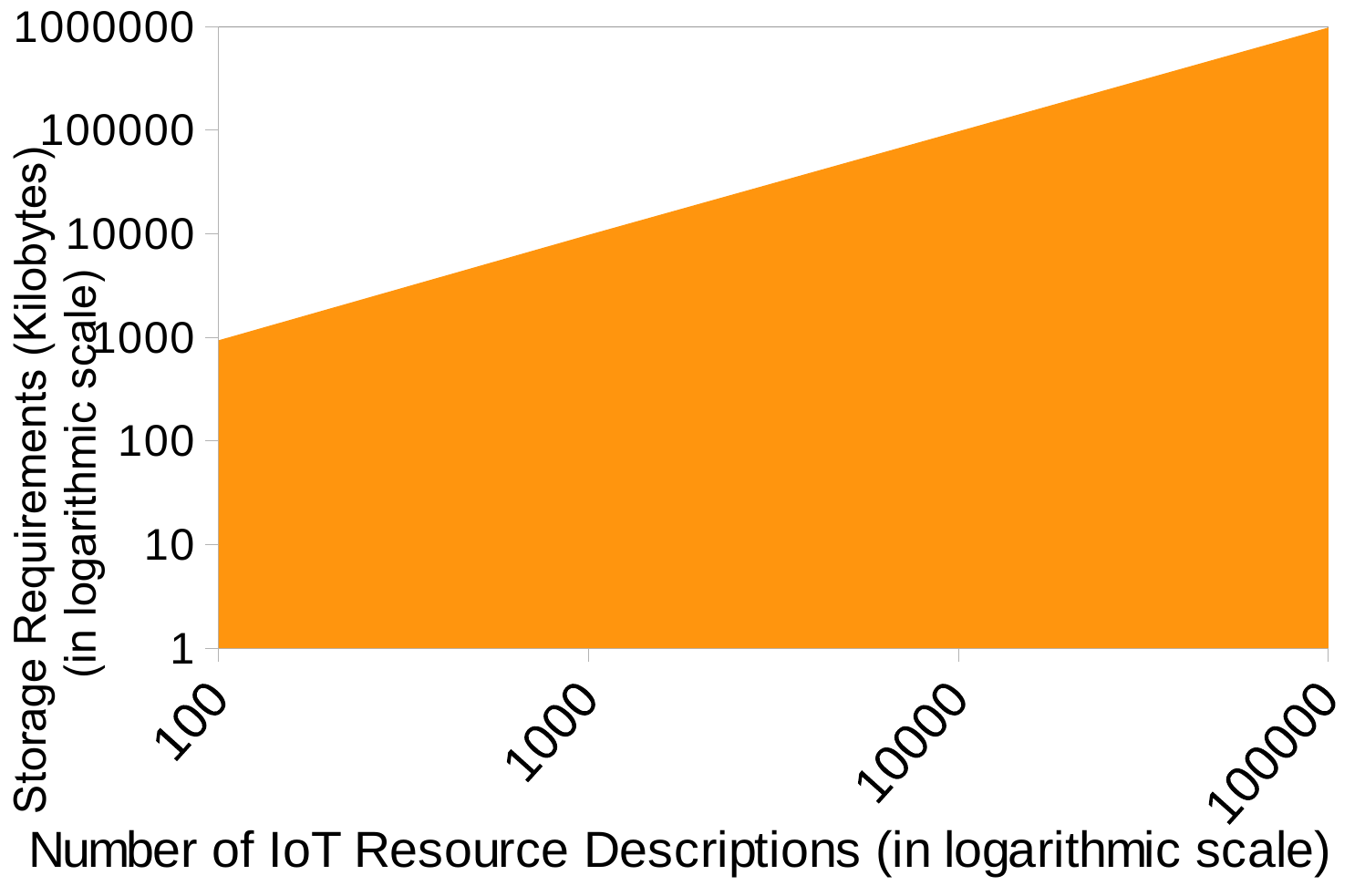}
                \vspace{-6pt}
                \caption{\footnotesize }
                \label{Figure:Results1_Storage_Requirment}
        \end{subfigure}%
        ~ 
        \begin{subfigure}[b]{165pt}
                \centering
                \includegraphics[scale=.34]{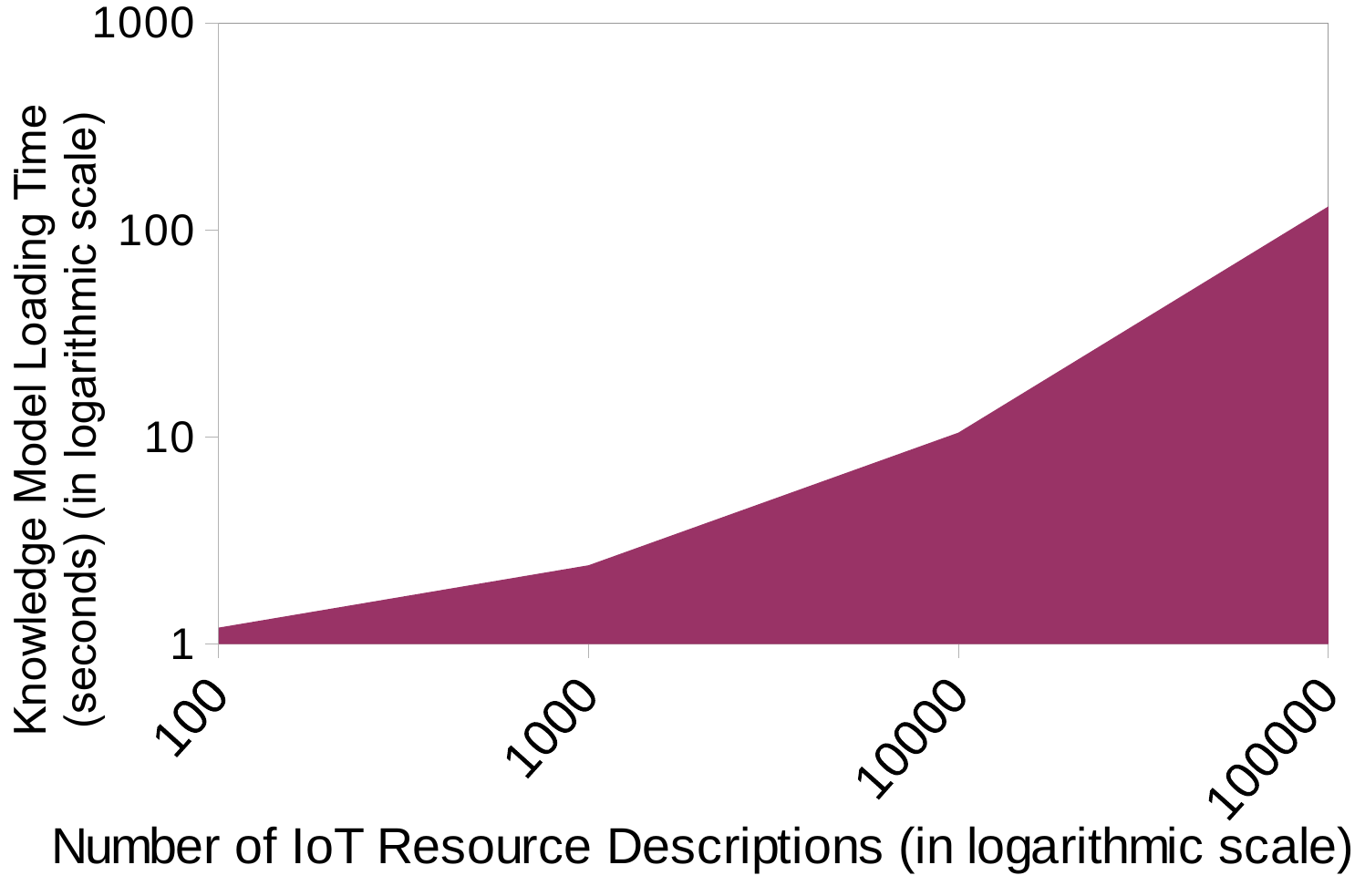}
                \vspace{-6pt}
                \caption{\footnotesize }
                \label{Figure:Results2_Loading_Time}
        \end{subfigure}
        ~ 
        \begin{subfigure}[b]{165pt}
                \centering
                \includegraphics[scale=.34]{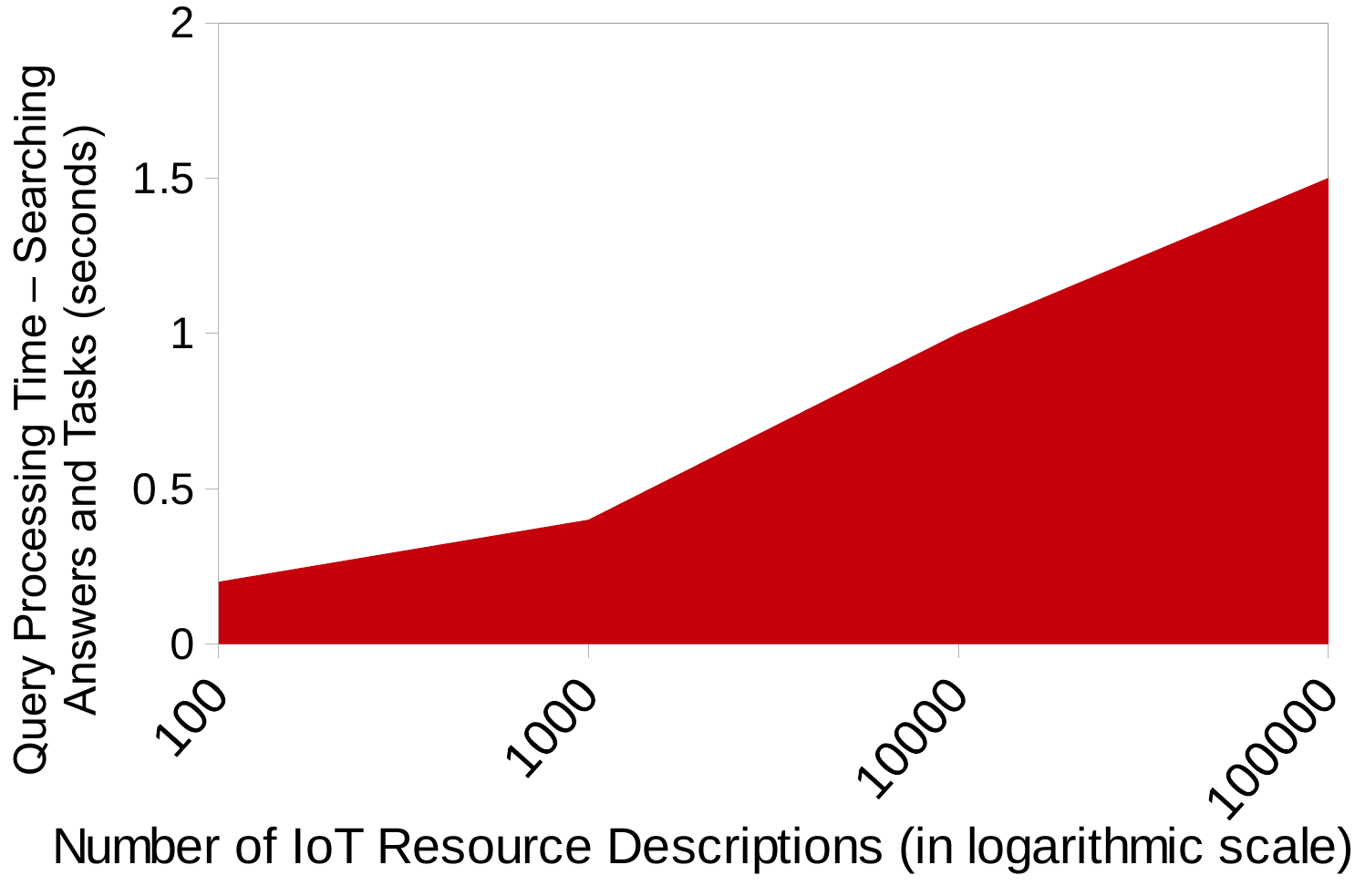}
                \vspace{-6pt}
                \caption{\footnotesize }
                \label{Figure:Results3_QA_Query_Time}
        \end{subfigure}

        \centering
        \begin{subfigure}[b]{165pt}
                \centering
                \includegraphics[scale=0.34]{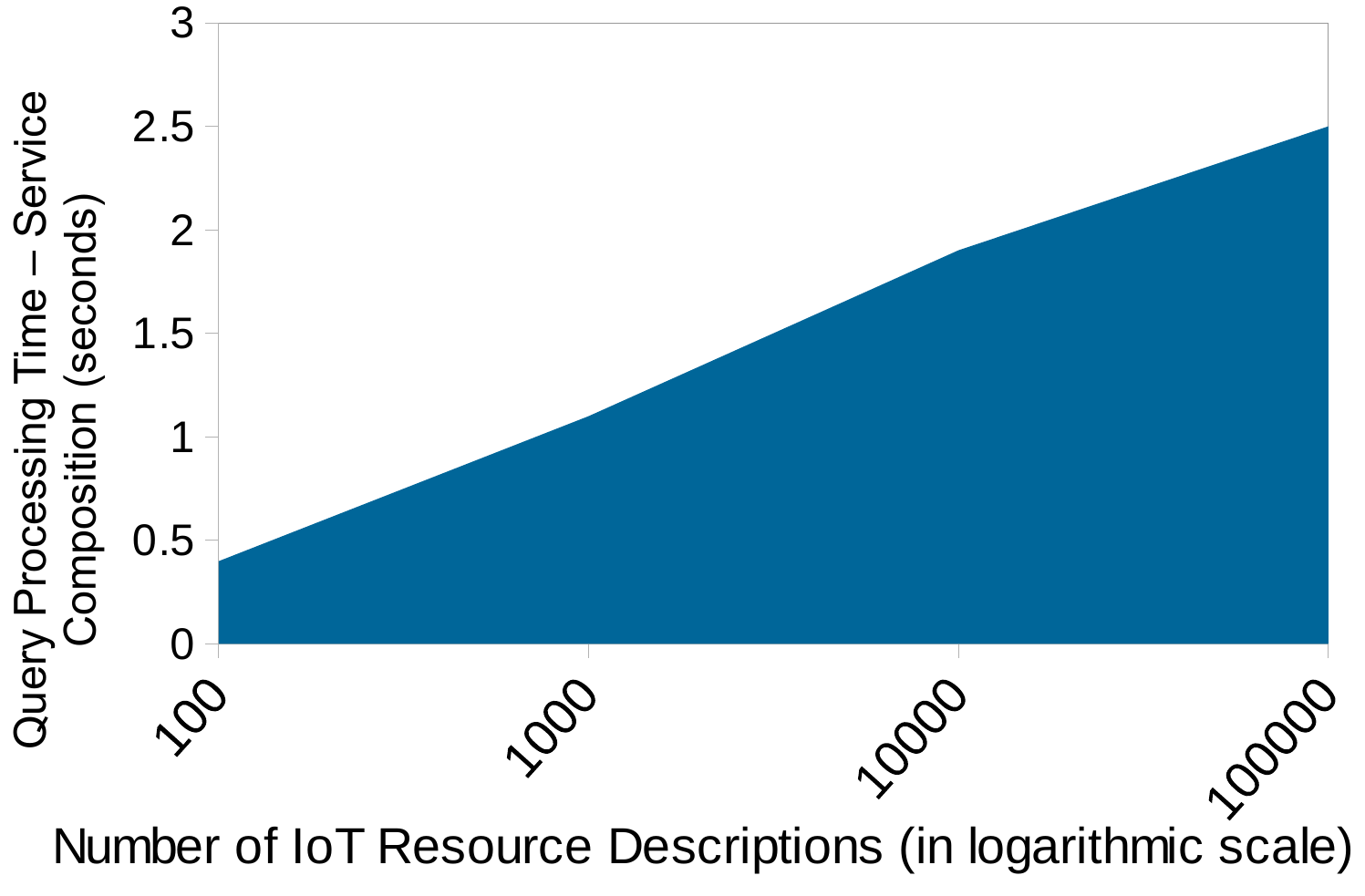}
                \vspace{-6pt}
                \caption{\footnotesize }
                \label{Figure:Results4_Composition_and_Context_Discovery_Time}
        \end{subfigure}%
        ~ 
        \begin{subfigure}[b]{165pt}
                \centering
                \includegraphics[scale=.34]{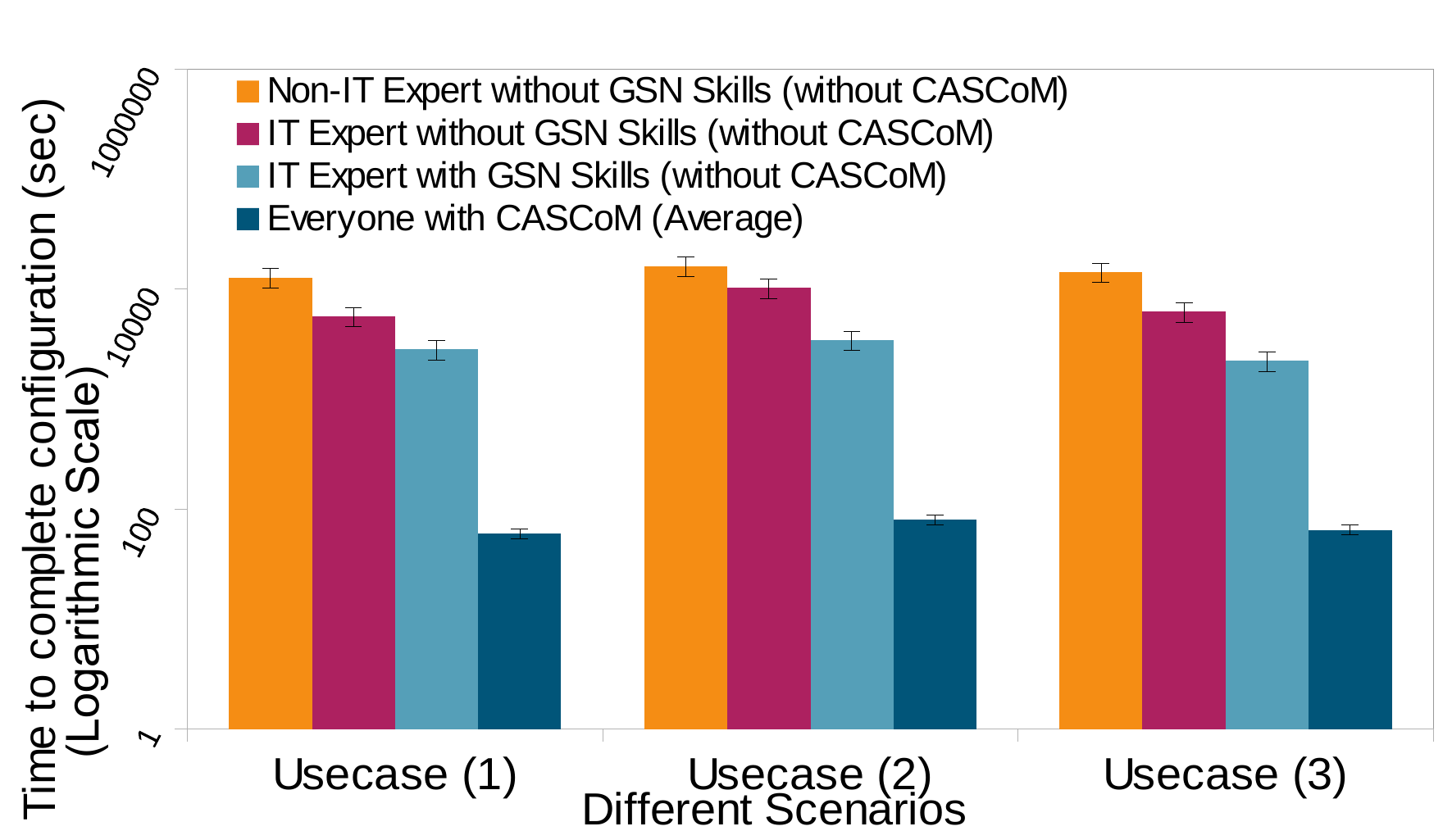}
                \vspace{-6pt}
                \caption{\footnotesize }
                \label{Figure:Results5_Usability}
        \end{subfigure}
        ~ 
        \begin{subfigure}[b]{165pt}
                \centering
                \includegraphics[scale=.34]{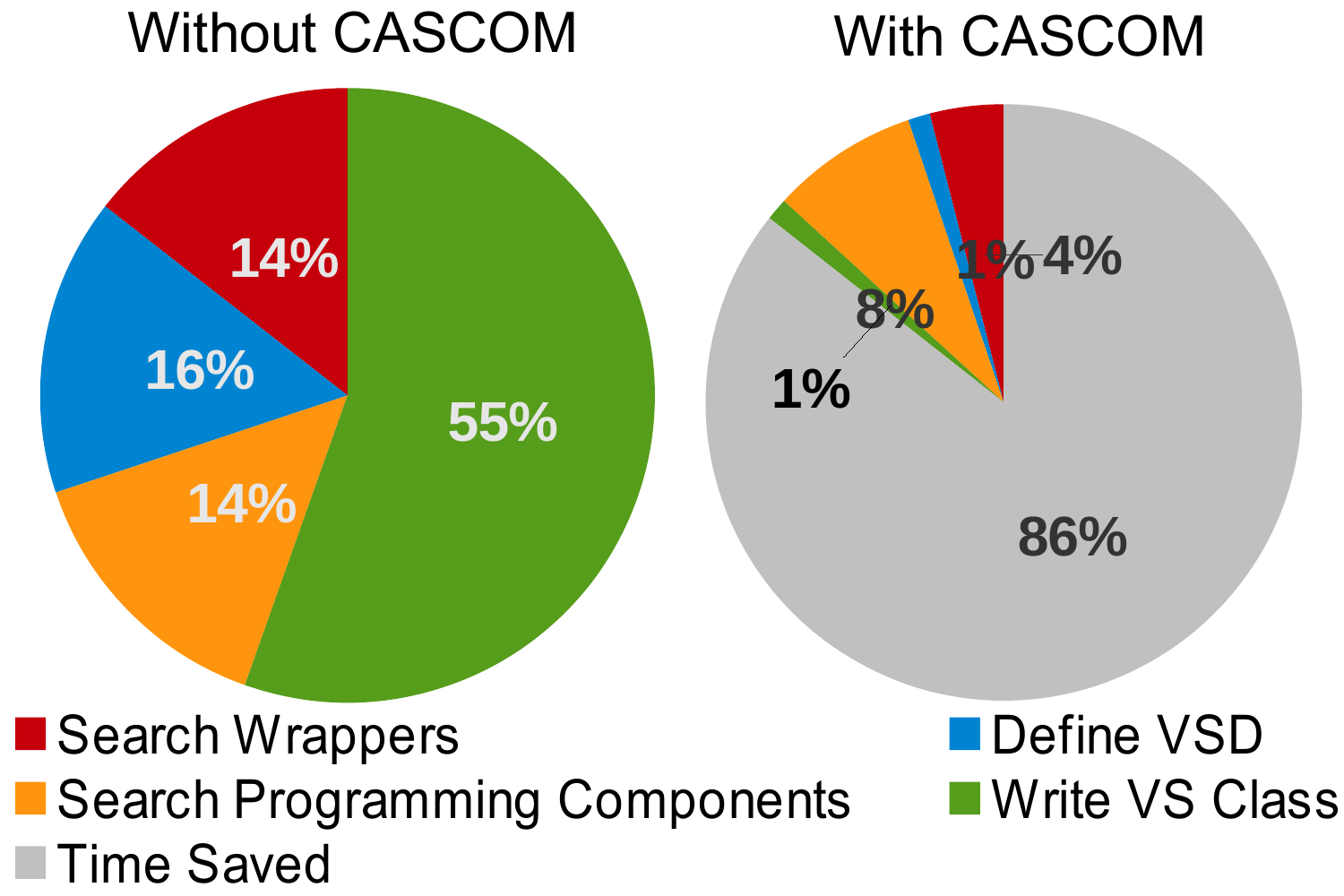}
                \vspace{-6pt}
                \caption{\footnotesize }
                \label{Figure:Results6_Time_Saving}
        \end{subfigure}
\vspace{-6pt}
\caption{CASCOM Performance Evaluations}
\vspace{-6pt}
\end{figure*}

 \section{Implementation and Experimentation}
 \label{sec:Implementation}

This section presents implementation details of our proof of concept development and evaluation from both computational and usability perspectives.

\subsection{Testbed}
\label{sec:AD:Testbed}

For proof of concept deployment and evaluation, we used a computer with Intel(R) Core i7 CPU and 16GB RAM. We used the Java programming language to develop the CASCOM tool and employed the open source Apache Jena API to manipulate semantic data. We used a Jena TDB-backed\footnote{jena.apache.org/documentation/tdb} approach to store the data. The user interface has been developed using the Java Swing framework. We modelled sensor descriptions according to the Semantic Sensor Network Ontology (SSN) \cite{P626}. Further, we modelled data processing components (DPC) descriptions according to the Software Component Ontology Plus (SCO+). The proposed SCO+ is based on SCO \cite{P612}, but additionally supports modelling context information such as accuracy and reliability as presented in Figure \ref{Figure:Ontology_Model}. 

To evaluate the proposed model, we developed a software tool that is illustrated in Figure \ref{Figure:User Interface1}. First, data consumers can select a question that they can answer from the drop down box. Then, they are allowed to answer the question. Possible answers will be listed in the next panel. Next, consumers can either answer another question by clicking \textit{Answer More} button. In contrast, they can click \textit{Search Tasks} button to search possible sensing tasks. Possible sensing tasks will be listed at the bottom of the next panel. Data consumers can select the sensing task they want and click \textit{Search Solution} button. CASCOM will  automatically generate different compositions of IoT resource that can perform the sensing task requested by the consumer.

\subsection{Methodology}
\label{sec:AD:Methodologybed}

We evaluated CASCOM using both qualitative and quantitative methods. We analysed and compared our proposed solution with respect to the existing GSN configuration model \cite{P022}. First, let us present our quantitative evaluation strategy (i.e. computational performance). In Figure \ref{Figure:Results1_Storage_Requirment}, we examined the feasibility of CASCOM model in term of how much data storage capacity is required as the knowledge-base grows. In Figure \ref{Figure:Results2_Loading_Time}, we examined the feasibility of CASCOM by measuring the variability of the data model loading time as the knowledge-base grows. Next, in Figure \ref{Figure:Results3_QA_Query_Time}, we evaluated the variability of query processing time, related to searching tasks, when the knowledge-base grows. In Figure \ref{Figure:Results4_Composition_and_Context_Discovery_Time}, we examined how IoT resource composition and secondary context discovery query processing time varied as the knowledge-base grows. In order to evaluate CASCOM's computational performance, we  generated a data model, according to the ontology presented in Figure \ref{Figure:Ontology_Model}, that consists of large amount of IoT resource descriptions. 

Now, let us present our qualitative evaluation strategy (i.e usability). We used three use-case scenarios for this evaluation. In each use case, the user was required to configure the IoT middleware in such a way that it produces a specific data stream: (1) \textit{monitor Phytophtora disease}, (2) \textit{monitor environmental pollution}, and (3) \textit{monitor and analyse crowd movement (indoor)}. Further, we selected three types of users: (1) \textit{IT experts who were familiar with GSN configuration process}, (2) \textit{IT experts who was were familiar with the GSN}, and (3) \textit{non-IT experts}.

For the usability study, we created a similar data model, but with a small number of IoT resource descriptions. Details of these data sets are presented below. For each use case, a set of basic instructions and programming guidelines that explains the GSN configuration process were given. First, we asked the users to configure the GSN middleware without the support of CASCOM. Secondly, we asked the users to configure the GSN middleware by using CASCOM. We measured the time taken by each user and results are presented in Figure \ref{Figure:Results5_Usability}. In this evaluation, we considered the time taken by both users as well as by the computer to perform resource selection and composition. Further, 31 participants (15 IT expert who were not familiar with the GSN, 15 non-IT experts, and 1 IT expert who was familiar with GSN configuration process) were involved in this experiment. In Figure \ref{Figure:Results6_Time_Saving}, we analysed different phases of the configuration process separately and compared the current approaches with the CASCOM approach. In order to make the results comparable, we assumed the users are IT experts who know the GSN configuration process.

\section{Results, Discussion and Lessons Learned}
\label{sec:Discussion}

As shown in Figure \ref{Figure:Results1_Storage_Requirment}, the storage requirement grows  linearly\footnote{Graphs in logarithmic scale.} when the knowledge-base grows. In semantic modelling, the data model loading time is proportionate to the data model size. Therefore, as expected, loading time also grows linearly when the knowledge-base grows as shown in Figure \ref{Figure:Results2_Loading_Time}. However, it is important to note that the actual data model size and the actual loading time vary based on the data modelling technique used (e.g. file-based, database-based) and the semantic framework employed \cite{P591} (e.g. Jena, Sesame). The amount of time that is required to load the CASCOM data model into memory is less than ~200 seconds even when it contains 100,000\footnote{100,000 means we have modelled 100,000 sensor descriptions, data processing components, and tasks related knowledge descriptions each.} descriptions. Similarly, Jena-TDB takes only ~1GB to model and store 100,000 IoT resource descriptions\footnote{In our synthetic data generation process, we assume each data processing component accepts three inputs and produce one output.}. In similar conditions, task searching query can return the results in less than 1.5 seconds as show in Figure \ref{Figure:Results3_QA_Query_Time}. Further, resource composition can also be completed in 2.5 seconds as shown in Figure \ref{Figure:Results4_Composition_and_Context_Discovery_Time}. When we consider real world deployments, it is very unlikely that a single instance os GSN middleware would host over a 100,000 sensors and data processing components connect to it. Based on these results we can conclude that CASCOM is feasible to use in real-world deployment.

In semantic data modelling, model size, storage requirement, and query times depend on the number of descriptions that are modelled in a given store. Let us consider the data model depicted in Figure \ref{Figure:Ontology_Model}. In this model (as well as in our simulations), we have used only a part of the SSN ontology, because the other parts are irrelevant for the composition process. However, if we want to model using the full SSNO, the model size would grow depending on how much more information (i.e. nodes and edges) that we want to include in order to describe a given set of IoT resources.

As also presented in Figure \ref{Figure:Results5_Usability}, non-IT experts required extremely detailed guidelines (compared to IT experts) to perform the configuration as they are not familiar with the activities such as programming. They also required direct verbal assistant from the authors. In addition, it was revealed that non-IT experts and IT experts who are not familiar with GSN were unable to configure the GSN at all without guidelines. In contrast, simple guidelines that explain the GUI allowed all users to complete the given task, using CASCOM, within a fairly similar amount of time. Though the complexity of the user requirement (i.e. configuration related to each scenario) makes visible impact on configuration time in the current GSN approach, it diminishes when users use CASCOM to configure GSN.  Figure \ref{Figure:Results5_Usability} shows that CASCOM allows to considerably reduce the time required for configuration of data processing mechanism in IoT middleware. Specifically, CASCOM allowed the three types of users to complete the given task approximately 40, 110 and 210 times faster (respectively) in comparison to the existing approach.

According to Figure \ref{Figure:Results6_Time_Saving}, even IT experts who know GSN can save time by using CASCOM up to 86\%. Specially, time taken for defining the Virtual Sensor Definition (VSD) and Virtual Sensor (VS) class have been significantly reduced \footnote{VSD, VS are both configuration files that need to be dealt with when configuring GSN middleware. More details are available in \cite{P022}}. Both files can be generated by CASCOM autonomously within a second even for complex scenarios. However, the time taken to find data processing components and sensors (and wrappers) depends on the size of the semantic data model.

As CASCOM models knowledge according to ontologies, users do not need to memorise domain knowledge (i.e. \textit{which sensor data types are required to perform a  certain task?}). This is an significant improvement over the existing approach. Due to the employment of semantic technologies, CASCOM is extensible into any domain. More importantly, adding new sensor descriptions and data processing component descriptions to the data model overtime allows CASCOM to compose new solutions. Ontological reasoning allows to deal with inconsistent usage of domain specific terminologies among domain experts. Ontologies helped in CASCOM to deal with performing validating task in composition of data components. Alternative to ontologies, we could have used a configuration file that explains which programming components and sensors need to be used to produce the required data stream for a given application (e.g. template-base approach). However, such an approach will drastically reduce the interoperability and flexibility. In IoT, ideal approaches should be able to dynamically compose and configure sensors and data processing components as it is impossible predict their availability at give time (new sensors and data processing components may available to use).

\subsection{Revisiting Challenges}

\textcolor{blue}{In this section, we summarise how the challenges and drawbacks identified  in the related work section are being addressed by our proposed solution. Main weakness in the existing solutions, such as Microsoft \textit{SensorMap} \cite{P578} and \textit{Linked Sensor Middleware} (LSM) \cite{P584},  is that they are user driven and not scalable and . Users are expected to conduct discovery and composition by themselves either using naked eye (i.e. looking at the user interface provided) or limited keyword-based search facilities. In contrast,  CASCOM is a knowledge driven approach where users only required to input very high level user requirement. The discovery and composition is done autonomously based on the knowledge model. As CASCOM models knowledge according to ontologies, users do not need to memorise domain knowledge. This is an significant improvement over the existing approach such as Microsoft \textit{SensorMap} and \textit{Linked Sensor Middleware} (LSM). Our results also show that knowledge driven approach allowed users to accomplish their task much faster than the user driven approaches.}

\textcolor{blue}{Additionally, we successfully demonstrated how hardware resources and software resources can be composed into work flows to achieve certain tasks. This is an advancement over existing approaches such as web service composition  \cite{P601, P613} where only software services are composed together.}

\section{Conclusions and Future Work}
\label{sec:Conclusions}

In this paper, we proposed a semantic  knowledge driven IoT resource discovery and composition engine to assist sensor data consumers to retrieve the data they want quickly and effortlessly. In particular, we focus on facilitating non-technical users to use IoT middleware platforms without spending too much time on learning technical details. To achieve this, we developed an IoT middleware configuration model called CASCOM. CASCOM makes the configuration process much easier by providing a sophisticate graphical user interface to express user requirements. Through a proof of concept implementation, we evaluated CASCOM both in term of usability and computational complexity. The results shows that the proposed model is significantly useful for non-technical personal to use IoT middleware platforms to retrieve data. CASCOM engine is highly flexible and scalable due to its knowledge driven nature where we can add more descriptions about data processing components and sensors over time. We have also done computational evaluations to demonstrate the feasibility and scalability of our proposed model. In additions to its primary role, CASCOM is capable of discovering secondary context through processing primary context information.

In the future, we would like to incorporate privacy aspects into the model. Currently, CASCOM is not considering any privacy violations that may occur when data processing components and sensors are composed together. It is important to evaluate and verify  all  consumer requests received by an IoT middleware to make sure that data owners' privacy are protected at all times. More importantly, for some consumer tasks (e.g. monitor and analyse crowd movement (indoor)), privacy would be a greater concern than for others (e.g. monitor \textit{Phytophtora} disease).

\section*{Acknowledgements}

Dr. Charith Perera's work has been funded by The Australian National University, The Commonwealth Scientific and Industrial Research Organisation (CSIRO), and European Research Council Advanced Grant 291652 (ASAP).

\bibliographystyle{elsart-num-sort}
\bibliography{Bibliography}

\vspace{20pt}

\setstretch{0.9}
\parpic{\includegraphics[width=1in,clip,keepaspectratio]{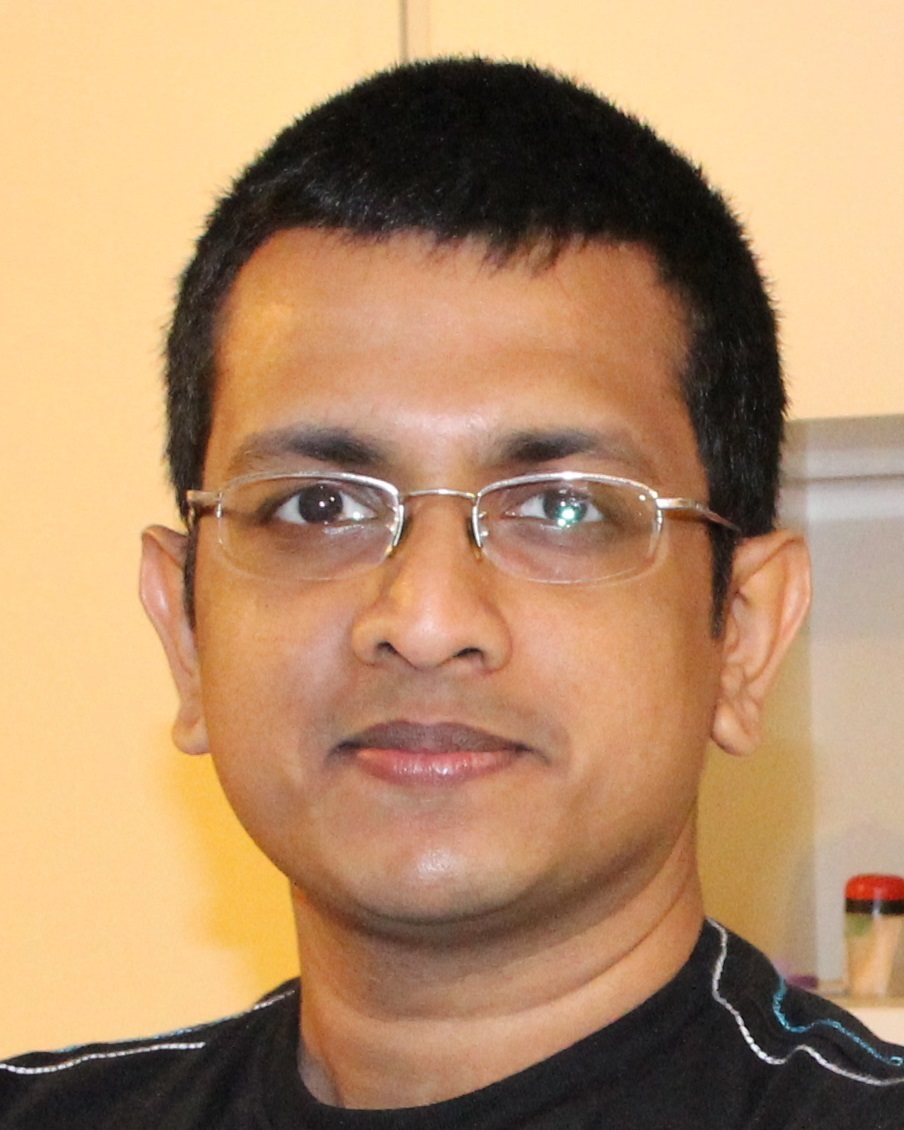}}
{ \small \noindent {\bf Charith Perera}  is a Research Associate  at The Open University, UK. Previously he worked at the Information Engineering Laboratory, ICT Center, Commonwealth Scientific and Industrial Research Organization (CSIRO). Perera received his BSc (Hons) in Computer Science  from Staffordshire University, Stoke-on-Trent, UK and MBA in Business Administration  from University of Wales, Cardiff, UK and PhD in Computer Science from The Australian National University, Canberra, Australia. His research interests include the Internet of Things, smart cities, mobile and pervasive computing, context-awareness, and ubiquitous computing. He is a member of both IEEE and ACM.  Contact him at charith.perera@ieee.org.}

\vspace{10pt}

\parpic{\includegraphics[width=1in,clip,keepaspectratio]{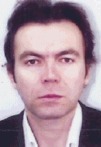}}
{ \small \noindent {\bf Athanasios V. Vasilakos }   is a professor at Dept of Computer Science, Electrical and Space Engineering, Lulea University of Technology, Lulea, Sweden. He has served as General Chair, and Techinical Program Committee Chair for many internation conferences. He also served or is serving as Editor or/and Guest Editor for many technical journals, such as IEEE TNSM, IEEE TSMC-partB, IEEE TITB, IEEE JSAC special issues of May 2009, Jan. 2011, March 2011, ACM TAAS and IEEE Communications Magazine. He is founding Editor-in-Chief of the International Journal of Adaptive and Autonomous Communications Systems (IJAACS) and the International Journal of Arts and Technology (IJART). He is also General Chair of the Council of Computing and Communications of the European Alliances for Innovation.}

\end{document}